\documentclass[epj]{svjour}
\usepackage{graphics}
\begin{document}




\title{Structural and decay properties of $Z=132,138$ superheavy nuclei}
\author{Asloob A. Rather\inst{1}, M. Ikram\inst{1}, A. A. Usmani\inst{1} Bharat Kumar\inst{2}, S. K. Patra \inst{2}
}                     
%
%
\institute{Department of Physics, Aligarh Muslim University, Aligarh-202002, India. \and Institute of Physics, Bhubaneswar-751 005, India.}
\date{Received: date / Revised version: date}
\abstract{
In this paper, we analyze the structural properties of $Z=132$ and $Z=138$ 
superheavy nuclei within the ambit of  axially deformed relativistic mean-field 
framework with NL$3^{*}$ parametrization and calculate the total binding energies, 
radii, quadrupole deformation parameter, separation energies, density distributions.
We also investigate the phenomenon of  shape coexistence by performing the 
calculations for prolate, oblate and spherical configurations. 
For clear presentation of nucleon distributions, the two-dimensional contour 
representation of individual nucleon density and total matter density has been made.
Further, a competition between possible decay modes such as 
$\alpha$-decay, $\beta$-decay and spontaneous fission of the isotopic chain of 
superheavy nuclei with $Z=132$ within the range 312 $\le$ A $\le$ 392 and 
318 $\le$ A $\le$ 398 for $Z=138$ is systematically analyzed 
within self-consistent relativistic mean field model.
From our analysis, we inferred that the $\alpha$-decay and spontaneous fission
 are the  principal modes of decay in majority of the isotopes of superheavy nuclei
 under  investigation apart from $\beta$ decay as dominant mode of decay in $^{318-322}138$ isotopes.
\PACS{
      {PACS-key}{discribing text of that key}   \and
      {PACS-key}{discribing text of that key}
     } 
} 
\maketitle

\section{Introduction}
The quest for searching the limits on nuclear mass and charge 
in superheavy valley, which is still a largely unexplored
area of research in nuclear physics, has been an intriguing endeavour
for nuclear physics community from past several decades. In otherwords,
the discovery of new elements with atomic number $Z>102$ in the 
laboratory is being pursued with great vigour nowadays.
The existence of superheavy nuclei (SHN) is the result of the interplay 
of the attractive nuclear force and the disruptive Coulomb 
repulsion between protons that favours fission. 
In principle, for SHN the shape of the classical nuclear 
droplet which is governed by surface tension
 and coulomb repulsion is unable to withstand the surface distortions
 making these nuclei susceptible to spontaneous fission. 
Thus, the stability of superheavy elements has 
become a longstanding fundamental nuclear science problem. 
Some of the topical issues that the nuclear physics community is 
looking to address in the  superheavy regime of the nuclear chart are: 
how a nucleus with a large atomic number, such as $Z=112$, survives 
the huge electrostatic repulsion between the protons, its 
physical and chemical properties, the extent of the superheavy 
region, i.e., to find an upper 
limit on the number of neutrons and protons that can be bound into one cluster,
and the existence of very long-lived superheavy nuclei.  
Theoretically, the mere existence of the heaviest elements 
with $Z>102$ is entirely due to quantal shell effects.
However, in the midsixties, with the invention of the shell-correction 
method, it was established that long-lived superheavy elements 
(SHE) with very large atomic numbers could exist due to the strong shell 
stabilization ~\cite{MS66,SGK66,M67,NTSSWGLMN69,MG69}.
By incorporating shell effects, it shall be quite interesting to 
explore the regions in (Z, N) plane where long-lived superheavy nuclei might be expected. 
Exploration of (Z, N) plane in superheavy valley is driven by the understanding of 
not only the nuclear structure but also the structure of stars 
and the evolution of universe. 
Pursuing this line of thought, the pioneering work on superheavy elements was
 performed in 1960s~\cite{MS66,M67,NTSSWGLMN69,MG69} and such studies 
were quite successful in reproducing the  already known half-lives by employing
 macroscopic-microscopic method (Nilsson-Strutinsky approach) with the 
folded-Yukawa deformed single-particle potential~\cite{moller} and with the 
Woods-Saxon deformed single-particle potential~\cite{CPDN83,SPC89,PS91}. 
Further, the outcome of these exhaustive investigations led to the 
understanding that the valley of superheavy nuclei is separated in proton and neutron number
from known heavy elements by a region of much higher instability.
In addition, several theoretical models which 
come under the aegis of macro-micro method like the fission 
model~\cite{PISG85}, cluster model~\cite{BMP92}, the 
density dependent M3Y(DDM3Y) effective model~\cite{B03}, the 
generalized liquid drop model (GLDM)~\cite{ZR07} etc and 
self-consistent models like the relativistic mean field (RMF) 
theory~\cite{SFM05}, Skyrme Hatree-Fock (SHF) model~\cite{PXLZ07} 
etc proved to be an effective tool for the successful description
of $\alpha$ decay from heavy and SHN. 

From the past three decades, the experimentalists have launched an expedition
for predicting the `island of superheavy elements', a region of 
increasing stable nuclei around $Z=114$, which has led to a burst of 
activity in the superheavy regime. 
The synthesis of SHN in laboratory is accomplished by fusion of heavy 
nuclei above the barrier ~\cite{HM00}. 
The two main processes employed for the synthesis of SHN are
cold fusion performed mainly at GSI, Darmstadt and  RIKEN Japan and 
hot fusion reactions performed at JINR-FLNR, Dubna. 
Until now, SHN with Z $\le$ 118 have been synthesized in the laboratory. 
The elements with Z = 110, 111 and 112 were produced in 
the experiments carried out at GSI ~\cite{H95,H97,SH95,H96,HM0}.
The fusion cross section was extremely small in production of 
$Z=112$ nucleus which led to the conclusion that the formation of further 
heavier elements would be very difficult by this process. 
The element with $Z=113$ was identified at RIKEN, 
Japan ~\cite{M4,KM4} using cold fusion reaction with a very low cross 
section $\sim$0.03 pb thus confirming the limitation of cold-fusion technique. 
The synthesis of $Z=113-118$ was performed 
successfully by the experimentalists from joint collaboration of JINR-FLNR, 
Dubna and Lawrence Liverpool National Laboratory along with an 
unsuccessful attempt on the production of $Z=120$ through hot fusion 
technique~\cite{YTO12,YTO01,YTO05,YTO0}. 
The isotopes of elements $Z=$112, 114, 116 and 118 were identified in 
fusion-evaporation reactions at low excitation energies by 
irradiation of $^{233,238}$U, $^{242}$Pu, $^{248}$Cm and $^{248}$Cf 
with $^{48}$Ca beams~\cite{OY7}. 
The element $Z=118$ and its immediate decay product, element with 
$Z=116$, were produced at Berkeley Lab's 88 inch cyclotron by 
bombarding targets of lead with an intense beam of high-energy krypton ions. 
The element $^{270}$Hs with $Z=108$ and $N=162$ was synthesized 
by Dvorak et al.~\cite{D6} by $^{26}$Mg + $^{248}$Cm reaction. 
Although the advancement in the accelerator facilities and the nuclear 
beam technologies have pushed the frontiers of nuclear chart especially 
in the superheavy region upto a great extent except for an 
attempt~\cite{O9} to produce $Z=120$ superheavy nuclei through the reaction 
$^{244}$Pu + $^{58}$Fe, there has been until now no evidence for the 
production of nuclei with $Z>118$. 
The short life times and the low production cross sections 
observed in fusion evaporation residues often increases the difficulty 
in synthesis of new superheavy nuclei and are posing a major difficulty 
to both theoreticians and experimentalists in understanding the various 
properties of superheavy nuclei.

Superheavy nuclei and their decay properties is one of the fastest growing
fields in nuclear science nowadays. The discovery of alpha decay by Becquerel 
in 1896 and subsequently the alpha theory of decay proposed by Gamow, 
Condon and Gurnay in 1928 has ushered a new era in nuclear science. 
Quantum mechanically, $\alpha$-decay occurs in heavy and superheavy nuclei by a 
tunnelling process through a coulomb barrier which is classically forbidden.
The alpha decay~\cite{PPG86,PPG06,PPGG06,SBC07,DLW07,ZR07} of the SHN is 
possible only if the shell effect supplies the extra binding energy and 
increases the barrier height of the fission. 
Thus, the beta stable nuclei with relatively longer half-life for 
spontaneous fission than that of alpha decay indicate that the dominant 
decay mode for such a superheavy nucleus might be alpha decay.
It is  worth mentioning here that the $\alpha$-decay is not the only 
mode of decay found in heavy nuclei but there is wealth of literature for 
$\beta$-decay, spontaneous fission (SF) and cluster decay also for such 
nuclei ~\cite{RJ84,HHP89,SPG80,SAPSGG06,QXLW09,PGG11,NGG11,PGG12,NGG12}.
Generally, alpha decay occurs in heavy and superheavy nuclei 
while as beta decay can occur throughout the periodic chart. 
The understanding of spontaneous fission and alpha decay on superheavy 
nuclei is rather more important than beta decay because the SHN with relative 
small alpha decay half-lives compared to SF half-lives will survive the 
fission and thus can be observed in the laboratory through alpha decay.
Hence, the $\alpha$-decay plays an indispensable role in the identification of 
new superheavy elements. 
Besides this, it has also been predicted that beta decay may play an important 
role for some of the superheavy nuclei~\cite{karpov2012}.
However, $\beta$-decay proceeds through a weak interaction, the process 
is slow and less favoured compared to SF and alpha decay.

It is worth mentioning that the alpha decay and spontaneous fission are the 
main decay modes for both heavy and superheavy nuclei with $Z>92$. 
Where, spontaneous fission acts as the limiting factor that 
decides the stability of superheavy nuclei and hence puts a limit 
on the number of chemical elements that can exist.
It was Bohr and Wheeler~\cite{BW39} in 1939 who predicted and described the 
mechanism of spontaneous fission process on the basis of liquid drop model 
and established a limit of $\frac{Z^{2}}{A}\approx48$, beyond which 
 nuclei are unstable against spontaneous fission, and later in 1940, 
Flerov et. al.~\cite{flerov1940} observed this phenomenon in $^{235}$U. 
This was followed by the several empirical formulas being proposed by 
various authors for calculating the half lives in spontaneous fission and 
the first attempt in this direction was made by Swiatecki~\cite{S55} who 
proposed a semi-empirical formula for spontaneous fission. 
Further, Ren et. al.~\cite{ren2005,xu2005} proposed a phenomenological 
formula for calculating the spontaneous fission half-lives, and recently 
Xu et. al.~\cite{xu2008} generalized an empirical formula for 
spontaneous fission half-lives of even-even nuclei.
Here, in present manuscript, within the structural studies we 
made an attempt to look for the competition among various 
possible modes of decay such as $\alpha$-decay, 
$\beta$-decay and SF of the isotopes of $Z=132$ and $Z=138$ superheavy 
elements with a neutron range 180 $\le$ N $\le$ 260 and predict 
the possible modes of decay.
The contents of the manuscript are organized as follows. 
The framework of relativistic mean-field formalism is outlined in section two.
The results and discussion is presented in section three.
Finally, section four contains the main summary and conclusions of this work. 

\section{Theoretical Formalism}
From last few decades, the RMF theory has achieved a great 
success in describing many of the nuclear phenomena. 
Over the non-relativistic case, it is quite better to reproduce 
the structural properties of nuclei throughout the periodic 
table~\cite{S92,GRT90,R96,SW86,BB77} near or far from the stability 
lines including superheavy region~\cite{sil2004}.
The starting point of the RMF theory is the
 basic Lagrangian containing nucleons interacting with $\sigma-$,
 $\omega-$ and $\rho-$meson fields. The photon field $A_{\mu}$ is 
included to take care of the Coulomb interaction of protons. 
The relativistic mean field Lagrangian density is 
expressed as~\cite{S92,GRT90,R96,SW86,BB77},

\begin{eqnarray}
{\cal L}&=&\bar{\psi_{i}}\{i\gamma^{\mu}
\partial_{\mu}-M\}\psi_{i}
+{\frac12}\partial^{\mu}\sigma\partial_{\mu}\sigma
-{\frac12}m_{\sigma}^{2}\sigma^{2}
-{\frac13}g_{2}\sigma^{3} \nonumber \\
&-&{\frac14}g_{3}\sigma^{4}-g_{s}\bar{\psi_{i}}\psi_{i}\sigma 
-{\frac14}\Omega^{\mu\nu}\Omega_{\mu\nu}+{\frac12}m_{w}^{2}V^{\mu}V_{\mu}\nonumber \\
&-&g_{w}\bar\psi_{i}\gamma^{\mu}\psi_{i}
V_{\mu}-{\frac14}\vec{B}^{\mu\nu}\vec{B}_{\mu\nu} 
+{\frac12}m_{\rho}^{2}{\vec{R}^{\mu}}{\vec{R}_{\mu}}-{\frac14}F^{\mu\nu}F_{\mu\nu} \nonumber\\
&-&g_{\rho}\bar\psi_{i}\gamma^{\mu}\vec{\tau}\psi_{i}\vec{R^{\mu}}-e\bar\psi_{i}
\gamma^{\mu}\frac{\left(1-\tau_{3i}\right)}{2}\psi_{i}A_{\mu} .
\end{eqnarray}
Here M, $m_{\sigma}$, $m_{\omega}$ and $m_{\rho}$ are the masses for nucleon, 
${\sigma}$-, ${\omega}$- and ${\rho}$-mesons and ${\psi}$ is its Dirac spinor. 
The field for the ${\sigma}$-meson is denoted by ${\sigma}$, ${\omega}$-meson 
by $V_{\mu}$ and ${\rho}$-meson by $R_{\mu}$. 
$g_s$, $g_{\omega}$, $g_{\rho}$ and $e^2/4{\pi}$=1/137 are the coupling 
constants for the ${\sigma}$, ${\omega}$, ${\rho}$-mesons and 
photon respectively.$g_2$ and $g_3$ are the self-interaction coupling constants for
 ${\sigma}$ mesons. By using the classical variational principle,
 we obtain the field equations for the nucleons and mesons. 
\begin{eqnarray}
\{-\bigtriangleup+m^2_\sigma\}\sigma^0(r_{\perp},z)&=&-g_\sigma\rho_s(r_{\perp},z)\nonumber\\
&-& g_2\sigma^2(r_{\perp},z)-g_3\sigma^3(r_{\perp},z) ,\\
\{-\bigtriangleup+m^2_\omega\}V^0(r_{\perp},z)&=&g_{\omega}\rho_v(r_{\perp},z) ,\\
\{-\bigtriangleup+m^2_\rho\}R^0(r_{\perp},z)&=&g_{\rho}\rho_3(r_{\perp},z) , \\
-\bigtriangleup A^0(r_{\perp},z)&=&e\rho_c(r_{\perp},z). 
\end{eqnarray}
The Dirac equation for the nucleons is written by
\begin{equation}
\{-i\alpha\bigtriangledown + V(r_{\perp},z)+\beta M^\dagger\}\psi_i=\epsilon_i\psi_i.
\end{equation}
The effective mass of the nucleon is
\begin{equation}
M^\dagger=M+S(r_{\perp},z)=M+g_\sigma\sigma(r_{\perp},z),
\end{equation}
and the vector potential is
\begin{equation}
 V(r_{\perp},z)=g_{\omega}V^{0}(r_{\perp},z)+g_{\rho}\tau_{3}R^{0}(r_{\perp},z)+
e\frac{(1-\tau_3)}{2}A^0(r_{\perp},z). 
\end{equation}
A static solution is obtained from the equations of motion to describe 
the ground state properties of nuclei.
The set of nonlinear coupled equations are solved self-consistently
 in an axially deformed harmonic oscillator basis $N_F=N_B=20$. 
The quadrupole deformation parameter $\beta_{2}$ is extracted from
 the calculated quadrupole moments of neutrons and protons through 
\begin{equation}
Q = Q_n + Q_p = \sqrt{\frac{16\pi}5} \left(\frac3{4\pi} AR^2\beta_2\right),
\end{equation}
where R = 1.2$A^{1/3}$.\\
The total energy of the system is given by 
\begin{equation}
 E_{total} = E_{part}+E_{\sigma}+E_{\omega}+E_{\rho}+E_{c}+E_{pair}+E_{c.m.},
\end{equation}
where $E_{part}$ is the sum of the single particle energies of the nucleons and 
$E_{\sigma}$, $E_{\omega}$, $E_{\rho}$, $E_{c}$, $E_{pair}$, $E_{cm}$ are 
the contributions of the meson fields, the Coulomb field, pairing energy  
and the center-of-mass energy, respectively. In present calculations, we use the 
constant gap BCS approximation to take care of pairing interaction~\cite{madland}.
We use non-linear NL3* parameter set~\cite{LKFAAR09} throughout the calculations. 

\section{Results and discussions}
In this paper, we performed self-consistent relativistic mean field 
calculations by employing NL$3^{*}$ for calculating the binding energy, radii 
and quadrupole deformation $\beta_{2}$ for three different shape configurations. 
In Refs.~\cite{zhang2005,mbhuyan}, $Z=132$, 138 are suggested to be 
proton and $N=198$, 228, 238 and 258 are neutron magic numbers.
Therefore, we considered a range of neutron $N=180-260$ that 
covers all these neutron magic numbers.
These neutron as well as proton magic numbers form the doubly magic systems 
as $^{330}132$, $^{360}132$, $^{370}132$, $^{366}138$, $^{376}138$ and $^{396}138$.
To analyze the structural properties of these isotopes, we made an attempt 
using deformed RMF calculations. 
It is well known that the superheavy nuclei are identified by $\alpha$-decay in the 
laboratory followed by spontaneous fission.
Therefore, to predict the possible mode of decay for the considered 
range of nuclides we make an investigation to analyze the 
competition between $\alpha$-decay, $\beta$-decay and spontaneous fission 
which is considered to be central theme of the paper.
The results are explained in subsections 3.1 to 3.6.

\subsection{Selection of basis space}
\begin{figure}
\vspace{0.30cm}
\resizebox{0.47\textwidth}{!}{
 \includegraphics{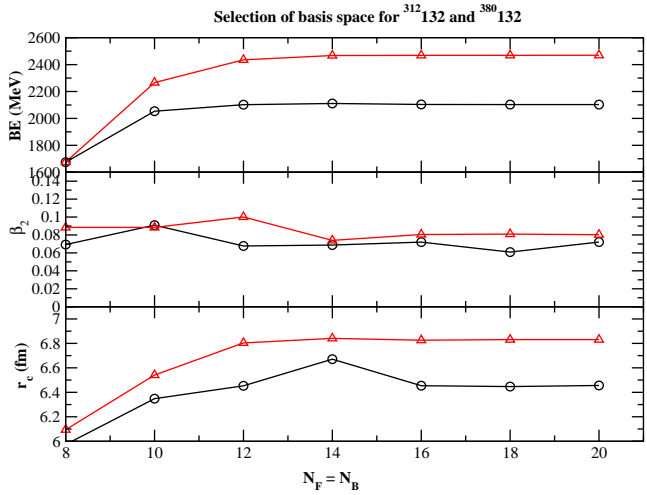}
}
\caption{(color online) The variation of calculated binding energy (BE), and 
quadrupole deformation parameter ($\beta_2$) and charge radius ($r_c$) 
are given with Bosonic and Fermionic basis.
Black lines with circle represents the results for $^{312}$132 and the 
outcome of $^{380}$132 are shown by red line with triangle.}
\label{basisspace}
\end{figure}

The RMF Lagrangian is used to obtain Dirac equation for Fermions and
the Klein-Gordon equations for Bosons using state-of the art 
variational approach in a self-consistent manner. 
Further, these equations are solved in an
axially deformed harmonic oscillator basis $N_F$ and
$N_B$ for Fermionic and Bosonic wavefunction, respectively.
For superheavy nuclei, a large number of basis space $N_F$ and $N_B$
is needed to get a convergent solution. 
For this, we have to choose an optimal model space
for both Fermion and Boson fields. 
To choose optimal values
for $N_F$ and $N_B$, we select $^{312,380}$132 systems as a test case 
and increase the basis number from 8 to 20 step by step. 
Results obtained for $^{312,380}$132 systems using these basis 
space are shown in Fig~\ref{basisspace}. 
From our calculations, we notice
an increment of 379 MeV in binding energy while going
from $N_F$ = $N_B$ = 8 to 10 for $^{312}132$ system and it comes 
to be 48 MeV while $N_{F}$ and $N_B$ changes from 10 to 12 and 
further by increasing the number of basis a constant value of BE is obtained. 
Proceeding along the similar lines, for $^{380}$132 system, we 
notice a large increment in binding energy around 590 MeV when 
the basis change form 8 to 10 and this amount of BE 
reduces to 170 MeV while the basis ($N_F$ and $N_B$) change 
form 10 to 12 and further a constant value of BE is obtained by 
increasing the basis space. 
This increment in energy decreases while going to higher oscillator basis. 
For example, change in binding energy is $\approx$0.2 and 0.6 MeV for 
$^{312}$132 and $^{380}$132 respectively with a change of $N_F$ = $N_B$ from 18 to 20.
Therefore, the present calculations dictate that the optimal basis 
sets to be chosen is $N_F$ = $N_B$ = 20 which is well within the 
convergence limits of the current RMF models.

\begin{figure}
\vspace{0.60cm}
\resizebox{0.47\textwidth}{!}{%
  \includegraphics{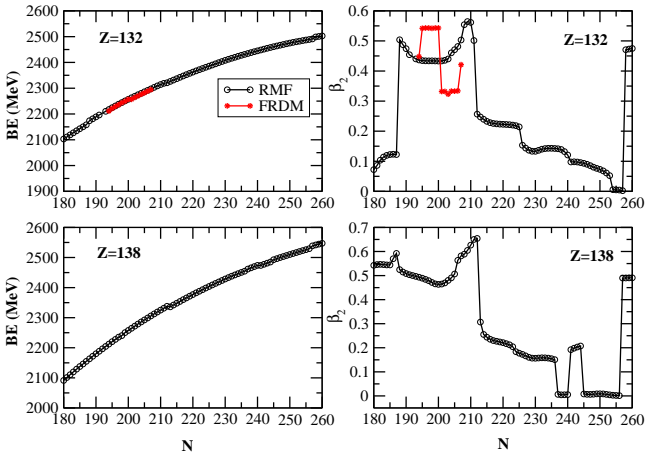}
}
\caption{(color online) Binding energy and deformation parameter as a function of neutron number.}
\label{binding}
\end{figure}

\subsection{Binding energy, radii and quadrupole deformation parameter}
\begin{figure}
\vspace{0.60cm}
\resizebox{0.47\textwidth}{!}{%
  \includegraphics{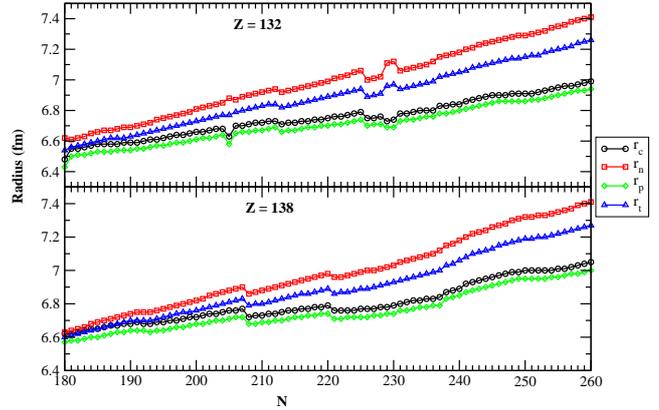}
}
\caption{(color online) Radii as a function of neutron number.}
\label{radii}
\end{figure}

The calculated binding energy, radii and the quadrupole deformation parameter 
for the isotopic chains $^{312-392}132$and $^{318-398}138$ are given
 in Tables~\ref{tab1} - \ref{tab4} and plotted in Figures~\ref{binding},~\ref{radii}. 
To find the ground state solution, the calculations are
 performed with an initial spherical, prolate and oblate quadrupole
 deformation parameter $\beta_{0}$ in the relativistic mean field formalism.
It is important to mention here that maximum binding 
energy corresponds to the ground state
 energy and all other solutions are the intrinsic excited state configurations. 
Proceeding along these lines, we found prolate as a ground state 
for most of the cases. As the experimental
 binding energies for these superheavy isotopic chains are not available,
in order to provide some validity to the predictive power of our calculations
 a comparison of binding energies of our calculations with those obtained
 from finite range droplet model (FRDM)~\cite{moller} is made 
wherever available and close agreement is found. 
The calculated quadrupole deformation parameter from RMF and the values 
obtained from FRDM~\cite{moller} predict the ground state of the considered 
isotopic chains to be prolate however there is a difference in magnitude 
as indicated in Table~\ref{tab1} as well as in Fig.~\ref{binding}. 
The radii monotonically increases with increasing number of neutrons. 
In general, the calculated binding energies are in good agreement 
with those of the FRDM values wherever available.

\begin{table*}
\caption{Binding energy, deformation parameter and radii for Z=132 
isotopic chain within three possible shape configurations.}
\renewcommand{\tabcolsep}{0.19cm}
\renewcommand{\arraystretch}{1.0}
\footnotesize
\begin{tabular}{c|ccc|ccc|ccc|ccc|cc}
\hline\hline
Nuclei&\multicolumn{3}{c|}{BE}&\multicolumn{3}{c|}{$\beta_2$}&\multicolumn{3}{c|}{$r_c$}&\multicolumn{3}{c|}{$r_t$}&\multicolumn{2}{c}{FRDM}\\
\cline{2-4} \cline{5-7} \cline{8-10} \cline{11-13} \cline{14-15}
&sph.&prol.&obl.&sph.&prol.&obl.&sph.&prol&obl.&sph.&prol.&obl.&BE&$\beta_2$\\
\hline
$^{312}132$&2103.0&	&	2159.2&	0.073&	&	-0.208&	6.457&	&	6.480&	6.479&	&	6.543&&\\
$^{313}132$&2111.0&	&	2118.0&	0.086&	&	-0.251&	6.469&	&	6.546&	6.493&	&	6.561&&\\
$^{314}132$&2119.0&	&	2126.0&	0.102&	&	-0.254&	6.483&	&	6.554&	6.508&	&	6.572&&\\
$^{315}132$&2127.0&	&	2133.8&	0.112&	&	-0.257&	6.494&	&	6.562&	6.521&	&	6.583&&\\
$^{316}132$&2134.9&	&	2141.4&	0.118&	&	-0.259&	6.503&	&	6.569&	6.532&	&	6.593&&\\
$^{317}132$&2142.7&	&	2148.9&	0.121&	&	-0.259&	6.511&	&	6.575&	6.543&	&	6.602&&\\
$^{318}132$&2150.4&	&	2156.2&	0.122&	&	-0.258&	6.519&	&	6.580&	6.553&	&	6.610&&\\
$^{319}132$&2157.9&	&	2163.5&	0.121&	&	-0.253&	6.525&	&	6.583&	6.562&	&	6.616&&\\
$^{320}132$&2165.2&	2173.8&	2170.8&	0.118&	0.496&	-0.245&	6.531&	6.753&	6.584&	6.571&	6.783&	6.620&&\\
$^{321}132$&2172.4&	2181.3&	2178.1&	0.113&	0.480&	-0.237&	6.535&	6.744&	6.585&	6.578&	6.778&	6.624&&\\
$^{322}132$&2179.6&	2188.7&	2185.4&	0.104&	0.467&	-0.233&	6.536&	6.739&	6.589&	6.584&	6.776&	6.631&&\\
$^{323}132$&2186.6&	2196.0&	2192.6&	0.091&	0.447&	-0.232&	6.535&	6.729&	6.594&	6.587&	6.769&	6.639&&\\
$^{324}132$&2193.7&	2196.0&	2199.7&	0.080&	0.447&	-0.232&	6.535&	6.729&	6.600&	6.591&	6.769&	6.648&&\\
$^{325}132$&2200.7&	2210.4&	2206.7&	0.072&	0.433&	-0.232&	6.538&	6.729&	6.607&	6.598&	6.775&	6.657&&\\
$^{326}132$&2207.7&	2217.4&	2213.6&	0.067&	0.429&	-0.233&	6.544&	6.732&	6.613&	6.606&	6.781&	6.667&2211.5&0.448\\
$^{327}132$&2214.6&	2224.4&	2220.3&	0.061&	0.425&	-0.234&	6.550&	6.736&	6.620&	6.615&	6.788&	6.677&2221.5&0.542\\
$^{328}132$&2221.3&	2231.1&	2226.9&	0.057&	0.420&	-0.236&	6.556&	6.741&	6.628&	6.624&	6.797&	6.687&2229.3&0.543\\
$^{329}132$&2227.8&	2237.8&	2233.4&	0.050&	0.421&	-0.238&	6.563&	6.747&	6.635&	6.633&	6.806&	6.697&2235.8&0.543\\
$^{330}132$&2234.3&	2244.3&	2239.7&	0.000&	0.428&	-0.239&	6.569&	6.754&	6.642&	6.641&	6.816&	6.707&2243.4&0.542\\
$^{331}132$&2240.7&	2250.7&	2245.9&	0.000&	0.428&	-0.241&	6.577&	6.761&	6.649&	6.651&	6.826&	6.718&2249.6&0.543\\
$^{332}132$&2246.7&	2256.8&	2252.1&	0.000&	0.426&	-0.242&	6.584&	6.768&	6.657&	6.661&	6.835&	6.728&2256.9&0.543\\
$^{333}132$&2252.5&	2262.9&	2258.2&	0.007&	0.427&	-0.244&	6.591&	6.776&	6.664&	6.671&	6.846&	6.738&2256.9&0.332\\
$^{334}132$&2258.4&	2268.7&	2264.2&	0.031&	0.429&	-0.245&	6.598&	6.785&	6.671&	6.680&	6.858&	6.748&2264.1&0.332\\
$^{335}132$&2264.4&	2274.5&	2270.1&	0.051&	0.434&	-0.246&	6.605&	6.795&	6.677&	6.690&	6.871&	6.758&2269.9&0.323\\
$^{336}132$&&	2280.1&	2275.9&&	0.455&		-0.248&	&	6.817&	6.684&	&	6.896&	6.767&2276.6&0.333\\
$^{337}132$&&	2285.9&	2294.1&&	0.469&		-0.186&	&	6.832&	6.627&	&	6.914&	6.771&2281.9&0.333\\
$^{338}132$&&	2291.6&	2287.2&&	0.480&		-0.251&	&	6.780&	6.697&	&	6.933&	6.787&2288.6&0.334\\
$^{339}132$&&	2297.2&	2292.7&&	0.494&		-0.253&	&	6.875&	6.703&	&	6.962&	6.797&2294.3&0.421\\
$^{340}132$&&	2303.2&	2298.1&&	0.514&		-0.255&	&	6.933&	6.710&	&	7.021&	6.807&&\\
$^{341}132$&&	2308.8&	2303.5&&	0.525&		-0.257&	&	6.949&	6.716&	&	7.040&	6.817&&\\
$^{342}132$&&	2314.1&	2308.7&&	0.516&		-0.259&	&	6.950&	6.722&	&	7.045&	6.827&&\\
$^{343}132$&&	2318.9&	2314.0&&	0.456&		-0.261&	&	6.898&	6.728&	&	6.997&	6.836&&\\
$^{344}132$&&	2321.0&	2319.1&&	0.258&		-0.260&	&	6.707&	6.733&	&	6.819&	6.844&&\\
$^{345}132$&&	2326.3&	2324.0&&	0.248&		-0.202&	&	6.707&	6.712&	&	6.823&	6.823&&\\
$^{346}132$&&	2331.6&	2329.2&&	0.240&		-0.201&	&	6.709&	6.717&	&	6.828&	6.831&&\\
$^{347}132$&&	2336.9&	2334.4&&	0.233&		-0.202&	&	6.713&	6.722&	&	6.834&	6.840&&\\
$^{348}132$&&	2342.2&	2339.4&&	0.229&		-0.203&	&	6.717&	6.727&	&	6.841&	6.849&&\\
$^{349}132$&&	2347.3&	2344.3&&	0.226&		-0.205&	&	6.721&	6.732&	&	6.849&	6.858&&\\
$^{350}132$&&	2352.4&	2349.2&&	0.225&		-0.207&	&	6.726&	6.738&	&	6.857&	6.867&&\\
$^{351}132$&&	2357.4&	2354.0&&	0.224&		-0.209&	&	6.731&	6.743&	&	6.866&	6.876&&\\
$^{352}132$&&	2362.3&	2358.7&&	0.223&		-0.212&	&	6.735&	6.749&	&	6.874&	6.886&&\\
$^{353}132$&&	2367.2&	2363.4&&	0.222&		-0.216&	&	6.740&	6.755&	&	6.883&	6.896&&\\
$^{354}132$&&	2371.8&	2368.1&&	0.222&		-0.221&	&	6.745&	6.761&	&	6.892&	6.906&&\\
$^{355}132$&2373.5&	2376.5&	2372.8&	-0.01&	0.219&	-0.227&	6.732&	6.750&	6.769&	6.868&	6.900&	6.918&&\\
$^{356}132$&&	2381.0&	2377.4&	&	0.217&	-0.237&	&	6.755&	6.780&	&	6.909&	6.932&&\\
$^{357}132$&&	2385.4&	2382.0&	&	0.214&	-0.242&	&	6.759&	6.789&	&	6.916&	6.944&&\\
$^{358}132$&&	2389.9&	&	&	0.155&	&	&	6.748&	&	&	6.906&	&&\\
$^{359}132$&2393.8&	2394.5&	&	0.001&	0.145&	&	6.752&	6.752&	&	6.901&	6.913&&&	\\
$^{360}132$&2398.6&	2399.1&	&	0.001&	0.138&	&	6.758&	6.757&	&	6.910&	6.921&&&	\\
$^{361}132$&2403.0&	2403.6&	2387.6&	0.001&	0.135&	-0.134&	6.764&	6.763&	6.734&	6.920&	6.929&	6.964&&\\
$^{362}132$&2407.1&	2407.9&	2391.0&	0.002&	0.135&	-0.139&	6.770&	6.768&	6.739&	6.929&	6.938&	6.974&&\\
$^{363}132$&2411.3	&2412.0	&	&0.005	&0.136	&	&6.777	&6.773	&	&6.939	&6.948	&&&\\
$^{364}132$&2415.4	&2416.0	&	&0.014	&0.140	&	&6.783	&6.777	&	&6.949	&6.957	&&&\\
$^{365}132$&2419.5	&2419.9	&	&0.023	&0.142	&	&6.790	&6.782	&	&6.959	&6.966	&&&\\
$^{366}132$&2423.6	&2423.7	&	&0.024	&0.143	&	&6.796	&6.786	&	&6.969	&6.976	&&&\\
$^{367}132$&2427.6	&2427.4	&	&0.013	&0.143	&	&6.803	&6.791	&	&6.979	&6.985	&&&\\
\hline\hline
\label{tab1}
\end{tabular}
\end{table*}


\begin{table*}
\caption{Table I is continued.}
\renewcommand{\tabcolsep}{0.27cm}
\renewcommand{\arraystretch}{1.0}
\footnotesize
\begin{tabular}{c|ccc|ccc|ccc|ccc}
\hline\hline
Nuclei&\multicolumn{3}{c|}{BE}&\multicolumn{3}{c|}{$\beta_2$}&\multicolumn{3}{c|}{$r_c$}&\multicolumn{3}{c}{$r_t$}\\
\cline{2-4} \cline{5-7} \cline{8-10} \cline{11-13}
&sph.&prol.&obl.&sph.&prol.&obl.&sph.&prol&obl.&sph.&prol.&obl.\\
\hline
$^{368}132$&2431.6	&2431.1	&2430.2	&0.004	&0.143	&-0.083	&6.81	&6.796	&6.803	&6.989	&6.994	&6.988\\
$^{369}132$&2435.5	&2434.6	&2433.4	&0.002	&0.142	&-0.172	&6.816	&6.8	&6.825	&6.999	&7.003	&7.015\\
$^{370}132$&2439.3	&2438.2	&2437.5	&0.001	&0.139	&-0.176	&6.822	&6.804	&6.831	&7.008	&7.011	&7.026\\
$^{371}132$&2442.6	&2441.5	&2441.5	&0.001	&0.132	&-0.179	&6.825	&6.806	&6.837	&7.017	&7.017	&7.036\\
$^{372}132$&2445.6	&2444.8	&2445.3	&0.002	&0.121	&-0.183	&6.828	&6.806	&6.843	&7.025	&7.022	&7.046\\
$^{373}132$&2448.5	&	&2448.8	&0.003	&	&-0.197	&6.83	&	&6.855	&7.033	&	&7.061\\
$^{374}132$&2451.4	&	&2452.4	&0.005	&	&-0.214	&6.831	&	&6.870	&7.041	&	&7.077\\
$^{375}132$&2454.3	&	&2455.9	&0.009	&	&-0.223	&6.833	&	&6.881	&7.049	&	&7.091\\
$^{376}132$&2457.2	&	&2459.2	&0.019	&	&-0.229	&6.834	&	&6.891	&7.056	&	&7.104\\
$^{377}132$&2460.8	&	&2462.2	&0.092	&	&-0.230	&6.821	&	&6.897	&7.058	&	&7.114\\
$^{378}132$&2463.8	&	&2464.6	&0.089	&	&-0.228	&6.824	&	&6.902	&7.066	&	&7.122\\
$^{379}132$&2466.7	&	&2467.6	&0.084	&	&-0.224	&6.827	&	&6.904	&7.074	&	&7.129\\
$^{380}132$&2469.6	&	&2470.3	&0.079	&	&-0.218	&6.831	&	&6.906	&7.083	&	&7.135\\
$^{381}132$&2472.3	&	&2473.1	&0.076	&	&-0.213	&6.834	&	&6.908	&7.091	&	&7.141\\
$^{382}132$&2475.1	&	&2475.8	&0.072	&	&-0.209	&6.837	&	&6.910	&7.099	&	&7.148\\
$^{383}132$&2477.6	&	&2478.6	&0.067	&	&-0.205	&6.839	&	&6.913	&7.107	&	&7.155\\
$^{384}132$&2479.9	&	&2481.3	&0.060	&	&-0.205	&6.841	&	&6.918	&7.115	&	&7.164\\
$^{385}132$&2482.3	&	&2483.9	&0.051	&	&-0.208	&6.842	&	&6.926	&7.123	&	&7.175\\
$^{386}132$&2484.7	&	&2486.5	&0.002	&	&-0.213	&6.853	&	&6.936	&7.137	&	&7.187\\
$^{387}132$&2486.9	&	&2488.9	&0.002	&	&-0.219	&6.853	&	&6.946	&7.144	&	&7.199\\
$^{388}132$&2488.9	&	&2491.5	&0.002	&	&-0.224	&6.852	&	&6.955	&7.151	&	&7.212\\
$^{389}132$&2491.2	&	&2494.0	&0.001	&	&-0.228	&6.851	&	&6.964	&7.158	&	&7.223\\
$^{390}132$&2493.2	&2498.8	&2496.4	&0.001	&0.471	&-0.231	&6.851	&7.132	&6.972	&7.165	&7.383	&7.235\\
$^{391}132$&2495.2	&2500.9	&2498.5	&0.000	&0.473	&-0.234	&6.857	&7.141	&6.980	&7.175	&7.395	&7.246\\
$^{392}132$&2496.8	&2502.9	&2500.4	&0.007	&0.475	&-0.237	&6.867	&7.149	&6.989	&7.186	&7.407	&7.257\\
$^{393}132$&2498.7	&2504.8	&2502.4	&0.029	&0.477	&-0.240	&6.879	&7.158	&6.998	&7.197	&7.419	&7.269\\
$^{394}132$&2500.8	&2506.7	&2504.3	&0.046	&0.479	&-0.243	&6.893	&7.167	&7.007	&7.210	&7.432	&7.281\\
$^{395}132$&2502.9	&2508.7	&2506.1	&0.056	&0.482	&-0.247	&6.907	&7.176	&7.016	&7.222	&7.444	&7.293\\
\hline\hline
\end{tabular}
\label{tab2}
\end{table*}


\begin{table*}
\caption{Same as Table 1 but Z=138 isotopic chain.}
\renewcommand{\tabcolsep}{0.27cm}
\renewcommand{\arraystretch}{1.0}
\footnotesize
\begin{tabular}{c|ccc|ccc|ccc|ccc}
\hline\hline
Nuclei&\multicolumn{3}{c|}{BE}&\multicolumn{3}{c|}{$\beta_2$}&\multicolumn{3}{c|}{$r_c$}&\multicolumn{3}{c}{$r_t$}\\
\cline{2-4} \cline{5-7} \cline{8-10} \cline{11-13}
&sph.&prol.&obl.&sph.&prol.&obl.&sph.&prol&obl.&sph.&prol.&obl.\\
\hline
$^{318}138$&2081.8&  	2091.4& 2087.4& 0.000&  0.542&  -0.260&	6.508&	6.809&	6.615&	6.506&	6.788&	6.600\\
$^{319}138$&2090.9&  	2101.0& 2096.8& 0.000&  0.546&  -0.263&	6.513&	6.817&	6.623&	6.513&	6.799&	6.611\\
$^{320}138$&2099.7&     2110.5& 2106.0& 0.000&  0.547&  -0.266& 6.517&	6.824&	6.631&	6.521&	6.809&	6.621\\
$^{321}138$&2108.5&     2119.8& 2114.9& 0.000&  0.546&  -0.269&	6.523&	6.829&	6.638&	6.529&	6.816&	6.631\\
$^{322}138$&2117.1&     2128.9& 2123.7& 0.001&  0.544&  -0.272& 6.528& 	6.833& 	6.646& 	6.538& 	6.823& 	6.642\\
$^{323}138$&2125.7&     2137.7& 2132.3& 0.001&  0.544&  -0.275& 6.534& 	6.838& 	6.653& 	6.547& 	6.831& 	6.652\\
$^{324}138$&2134.2&     2146.2& 2140.7& 0.002&  0.570&  -0.277& 6.540& 	6.868& 	6.660& 	6.556& 	6.863& 	6.662\\
$^{325}138$&2142.6&     2154.8& 2149.0& 0.002&  0.592&  -0.279& 6.547& 	6.896& 	6.667& 	6.565& 	6.892& 	6.672\\
$^{326}138$&2151.1&     2163.3& 2157.2& 0.003&  0.525&  -0.280& 6.554& 	6.837& 	6.673& 	6.574& 	6.840& 	6.681\\
$^{327}138$&2159.5&     2171.7& 2165.3& 0.004&  0.515&  -0.280& 6.561& 	6.834& 	6.679& 	6.584& 	6.840& 	6.690\\
$^{328}138$&2167.9&     2180.0& 2173.2& 0.005&  0.508&  -0.278& 6.568& 	6.834& 	6.684& 	6.593& 	6.843& 	6.698\\
$^{329}138$&2176.2&     2188.3& 2181.0& 0.003&  0.504&  -0.272& 6.575& 	6.836& 	6.686& 	6.602& 	6.848& 	6.702\\
$^{330}138$&2184.5&     2196.4& 2188.9& 0.002&  0.500&  -0.260& 6.582& 	6.839& 	6.683& 	6.612& 	6.854& 	6.702\\
$^{331}138$&2192.9&     2204.5& 2196.8& 0.001&  0.497&  -0.249& 6.589& 	6.842& 	6.682& 	6.621& 	6.860& 	6.704\\			 
$^{332}138$&2201.1&     2212.4& 2204.7& 0.001&  0.493&  -0.245& 6.596& 	6.845& 	6.685& 	6.631& 	6.866& 	6.709\\
$^{333}138$&2209.4&     2220.2& 2212.4& 0.000&  0.489&  -0.243& 6.603& 	6.848& 	6.690& 	6.640& 	6.871& 	6.717\\
$^{334}138$&2217.6&   	2227.8& 2220.1& 0.000&  0.484&  -0.244& 6.610& 	6.850& 	6.696& 	6.649& 	6.876& 	6.726\\
$^{335}138$&2225.6&   	2235.4& 2227.6& 0.000&  0.479&  -0.244& 6.616& 	6.851& 	6.703& 	6.658& 	6.881& 	6.735\\
$^{336}138$&2233.3&   	2240.4& 2235.0& 0.000&  0.471&  -0.245& 6.622& 	6.851& 	6.710& 	6.667& 	6.883& 	6.745\\
$^{337}138$&2240.7&   	2250.3& 2242.3& 0.000&  0.465&  -0.247& 6.628& 	6.853& 	6.716& 	6.676& 	6.887& 	6.754\\
$^{338}138$&2247.7& 	2257.6& 2249.5& 0.000&  0.463&  -0.248& 6.634& 	6.858& 	6.723& 	6.684& 	6.895& 	6.764\\
$^{339}138$&2254.3&   	2264.8& 2256.7& 0.000&  0.465&  -0.248& 6.639& 	6.867& 	6.730& 	6.693& 	6.907& 	6.773\\
$^{340}138$&2261.1&   	2272.0& 2263.7& 0.000&  0.470&  -0.249& 6.645& 	6.879& 	6.736& 	6.701& 	6.921& 	6.782\\
$^{341}138$&2267.6&   	2279.0& 2270.7& 0.001&  0.480&  -0.250& 6.650& 	6.896& 	6.743& 	6.709& 	6.942& 	6.791\\
$^{342}138$&2274.0&  	2285.9& 2277.0& 0.001&  0.491&  -0.251& 6.656& 	6.914& 	6.749& 	6.717& 	6.962& 	6.801\\
$^{343}138$&2280.5&   	2292.5& 2284.1& 0.003&  0.506&  -0.252& 6.661& 	6.936& 	6.756& 	6.726& 	6.986& 	6.810\\
$^{344}138$&2287.0&     2299.1& 2290.6& 0.010&  0.563&  -0.254& 6.667& 	6.995& 	6.762& 	6.734& 	7.046& 	6.820\\
$^{345}138$&2293.4&   	2305.8& 2297.0& 0.025&  0.582&  -0.256& 6.674& 	7.020& 	6.769& 	6.743& 	7.072& 	6.829\\
$^{346}138$&2300.0& 	2312.5& 2301.8& 0.038&  0.589&  -0.158& 6.681& 	7.033& 	6.722& 	6.753& 	7.089& 	6.787\\
$^{347}138$&2306.4&     2319.0& 2308.4& 0.050&  0.604&  -0.157& 6.690& 	7.054& 	6.728& 	6.763& 	7.112& 	6.795\\
$^{348}138$&2312.9&   	2325.2& 2314.8& 0.095&  0.627&  -0.157& 6.706& 	7.078& 	6.733& 	6.780& 	7.139& 	6.803\\
$^{349}138$&&   	2331.4& 2321.0& &  0.650&  -0.157& & 	7.103& 	6.739& 	& 	7.166& 	6.812\\
$^{350}138$&&   	2337.9& 2327.1& &  0.654&  -0.157& & 	7.113& 	6.744& 	& 	7.180& 	6.820\\
$^{351}138$&& 	2336.0& 2333.2& &0.307  &  -0.158& & 	6.800& 	6.750& 	& 	6.879& 	6.829\\
$^{352}138$&& 	2342.1& 2339.0& &0.255  &  -0.159& & 	6.771& 	6.756& 	& 	6.857& 	6.838\\
$^{353}138$&& 	2348.1& 2344.9& &0.244  &  -0.161& & 	6.772& 	6.762& 	& 	6.860& 	6.847\\
$^{354}138$&& 	2354.0& 2350.7& &0.235  &  -0.162& & 	6.774& 	6.768& 	& 	6.865& 	6.856\\
$^{355}138$&&   	2359.9& 2356.4& &  0.230&  -0.163& & 	6.778& 	6.774& 	& 	6.872& 	6.864\\
$^{356}138$&& 	2365.7& 2361.9& &  0.227&  -0.163& & 	6.782& 	6.779& 	& 	6.879& 	6.873\\
$^{357}138$&& 	2371.4& 2367.4& &  0.224&  -0.163& & 	6.787& 	6.783& 	& 	6.887& 	6.881\\
$^{358}138$&& 	2377.0& 2372.9& &  0.221&  -0.161& & 	6.791& 	6.787& 	& 	6.894& 	6.888\\
$^{359}138$&& 	2382.5& 2380.0& &  0.217&  -0.158& & 	6.795& 	6.757& 	& 	6.902& 	6.862\\
$^{360}138$&& 	2387.9& 2385.8& &  0.212&  -0.151& & 	6.800& 	6.760& 	& 	6.909& 	6.868\\
$^{361}138$&2391.5& 	2393.2& & 0.019&  0.204&  & 6.758& 	6.803& 	& 	6.871& 	6.915& 	\\
$^{362}138$&2397.2& 	2398.6& & 0.005&  0.183&  & 6.761& 	6.804& 	& 	6.877& 	6.918& 	\\
$^{363}138$&2403.0& 	2403.7& & 0.003&  0.178&  & 6.765& 	6.809& 	& 	6.885& 	6.924& 	\\
$^{364}138$&2408.8& 	2408.8& & 0.002&  0.173&  & 6.770& 	6.813& 	& 	6.892& 	6.932& 	\\
$^{365}138$&2414.4& 	2413.8& & 0.001&  0.166&  & 6.774& 	6.816& 	& 	6.900& 	6.938& 	\\
$^{366}138$&2419.7& 	2418.8& & 0.001&  0.160&  & 6.779& 	6.820& 	& 	6.908& 	6.945& 	\\
$^{367}138$&2424.8& 	2423.8& & 0.001&  0.157&  & 6.784& 	6.825& 	& 	6.917& 	6.953& 	\\
$^{368}138$&2429.6& 	2428.6& & 0.002&  0.156&  & 6.790& 	6.830& 	& 	6.926& 	6.961& 	\\
$^{369}138$&2434.2	&2433.3	&	&0.003	&0.157	&	&6.795	&6.835	&	&6.935	&6.970	&\\
$^{370}138$&2438.8	&2437.9	&	&0.005	&0.158	&	&6.801	&6.839	&	&6.944	&6.978	&\\
$^{371}138$&2443.5	&2442.3	&	&0.008	&0.158	&	&6.807	&6.844	&	&6.954	&6.987	&\\
$^{372}138$&2448.0	&2446.4	&	&0.009	&0.157	&	&6.813	&6.848	&	&6.963	&6.995	&\\
$^{373}138$&2452.6	&2450.5	&	&0.006	&0.155	&	&6.818	&6.852	&	&6.972	&7.003	&\\
\hline\hline
\end{tabular}
\label{tab3}
\end{table*}


\begin{table*}
\caption{Table III is continued.}
\renewcommand{\tabcolsep}{0.27cm}
\renewcommand{\arraystretch}{1.0}
\footnotesize
\begin{tabular}{c|ccc|ccc|ccc|ccc}
\hline\hline
Nuclei&\multicolumn{3}{c|}{BE}&\multicolumn{3}{c|}{$\beta_2$}&\multicolumn{3}{c|}{$r_c$}&\multicolumn{3}{c}{$r_t$}\\
\cline{2-4} \cline{5-7} \cline{8-10} \cline{11-13}
&sph.&prol.&obl.&sph.&prol.&obl.&sph.&prol&obl.&sph.&prol.&obl.\\
\hline
$^{374}138$&2457.1	&2454.5	&	&0.004	&0.151	&	&6.824	&6.855	&	&6.981	&7.010	&\\
$^{375}138$&2461.6	&	&	&0.002	&	&	&6.829	&	&	&6.989	&	&\\
$^{376}138$&2465.9	&	&2465.7	&0.002	&	&-0.150	&6.833	&	&6.874	&6.998	&	&7.032\\
$^{377}138$&2469.9	&	&2470.2	&0.002	&	&-0.159	&6.837	&	&6.883	&7.006	&	&7.044\\
$^{378}138$&2473.9	&	&2474.7	&0.002	&	&-0.167	&6.839	&	&6.892	&7.014	&	&7.056\\
$^{379}138$&2477.7	&2473.7	&2478.9	&0.003	&0.192	&-0.203	&6.841	&6.908	&6.918	&7.021	&7.075	&7.083\\
$^{380}138$&2481.5	&2477.5	&2483.3	&0.004	&0.198	&-0.214	&6.843	&6.919	&6.930	&7.028	&7.087	&7.097\\
$^{381}138$&2485.3	&2481.3	&2487.7	&0.005	&0.203	&-0.222	&6.845	&6.929	&6.940	&7.036	&7.100	&7.110\\
$^{382}138$&2488.9	&2484.9	&2491.8	&0.005	&0.207	&-0.226	&6.847	&6.938	&6.949	&7.043	&7.111	&7.122\\
$^{383}138$&2492.6	&	&2495.5	&0.003	&	&-0.229	&6.849	&	&6.957	&7.051	&	&7.133\\
$^{384}138$&2496.3	&	&2499.0	&0.003	&	&-0.235	&6.852	&	&6.965	&7.058	&	&7.145\\
$^{385}138$&2499.8	&	&2502.5	&0.003	&	&-0.241	&6.854	&	&6.975	&7.066	&	&7.157\\
$^{386}138$&2503.3	&	&2505.9	&0.004	&	&-0.246	&6.857	&	&6.985	&7.075	&	&7.169\\
$^{387}138$&2506.7	&	&2509.2	&0.004	&	&-0.248	&6.860	&	&6.992	&7.083	&	&7.180\\
$^{388}138$&2509.9	&	&2512.4	&0.006	&	&-0.246	&6.863	&	&6.996	&7.091	&	&7.187\\
$^{389}138$&2513.4	&	&2515.5	&0.005	&	&-0.239	&6.866	&	&6.997	&7.100	&	&7.192\\
$^{390}138$&2516.6	&	&2518.8	&0.004	&	&-0.231	&6.868	&	&6.998	&7.108	&	&7.196\\
$^{391}138$&2519.9	&	&2521.9	&0.003	&	&-0.225	&6.871	&	&6.999	&7.117	&	&7.201\\
$^{392}138$&2523.0	&	&2525.3	&0.001	&	&-0.222	&6.873	&	&7.002	&7.125	&	&7.208\\
$^{393}138$&2526.3	&	&2528.6	&0.001	&	&-0.221	&6.876	&	&7.007	&7.134	&	&7.216\\
$^{394}138$&2529.4	&	&2531.8	&0.001	&	&-0.223	&6.879	&	&7.014	&7.142	&	&7.226\\
$^{395}138$&2532.4	&2538.7	&2534.9	&0.000	&0.489	&-0.226	&6.882	&7.194	&7.021	&7.151	&7.399	&7.236\\
$^{396}138$&2535.3	&2541.7	&2537.8	&0.000	&0.490	&-0.229	&6.887	&7.201	&7.029	&7.160	&7.409	&7.247\\
$^{397}138$&2537.9	&2544.6	&2540.7	&0.000	&0.490	&-0.232	&6.893	&7.207	&7.037	&7.170	&7.419	&7.258\\
$^{398}138$&2540.4	&2547.3	&2543.5	&0.000	&0.491	&-0.235	&6.901	&7.214	&7.045	&7.180	&7.429	&7.269\\
$^{399}138$&2542.7	&2550.2	&2546.2	&0.001	&0.491	&-0.238	&6.910	&7.222	&7.054	&7.190	&7.439	&7.281\\
$^{400}138$&2545.1	&2553.1	&2548.8	&0.001	&0.492	&-0.241	&6.919	&7.229	&7.063	&7.200	&7.450	&7.292\\
$^{401}138$&2547.4	&2555.9	&2551.4	&0.003	&0.494	&-0.245	&6.928	&7.236	&7.072	&7.210	&7.461	&7.303\\
\hline\hline
\end{tabular}
\label{tab4}
\end{table*}

\begin{figure}
\vspace{0.5cm}
\resizebox{0.47\textwidth}{!}{%
  \includegraphics{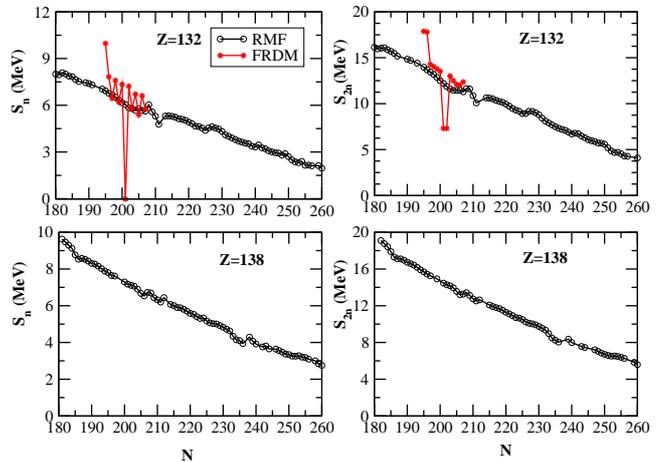}
}
\caption{(color online) One and two neutron separation energies as function of N}
\label{nsep}
\end{figure}

\subsection{Separation energy}
The separation energy is an important observable in identifying the
signature of magic numbers in nuclei. 
The magic numbers in nuclei are characterized 
by large shell gaps in their single-particle energy levels. 
This implies that the nucleons occupying the lower energy level have 
comparatively large value of energy than those nucleons occupying the 
higher energy levels, giving rise to more stability. 
The extra stability attributed to certain numbers can 
be predicted from the sudden fall in neutron separation energy. 
Two-neutron separation energy is more interesting which 
takes care of even-odd effects.
The one and two-neutron separation energy is calculated by the 
difference in binding energies of two isotopes using the relations
\begin{eqnarray}
S_{n}(N,Z) &= E_{B}(N,Z) - E_{B}(N-1,Z)\nonumber \\
S_{2n}(N,Z)&= E_{B}(N,Z) - E_{B}(N-2,Z)
\end{eqnarray}                                                                   
One- and two-neutron separation energy ($S_{n}$ and $S_{2n}$) for
 the considered isotopic series of the nuclei $^{312-392}$132 and 
$^{318-398}$138 are shown in Figure~\ref{nsep}.
No sudden fall of the separation energies is observed for both the cases which 
indicates that as such no neutron magic behaviour within this force parameter is noticed.

\begin{figure}
\vspace{0.4cm}
\resizebox{0.47\textwidth}{!}{%
  \includegraphics{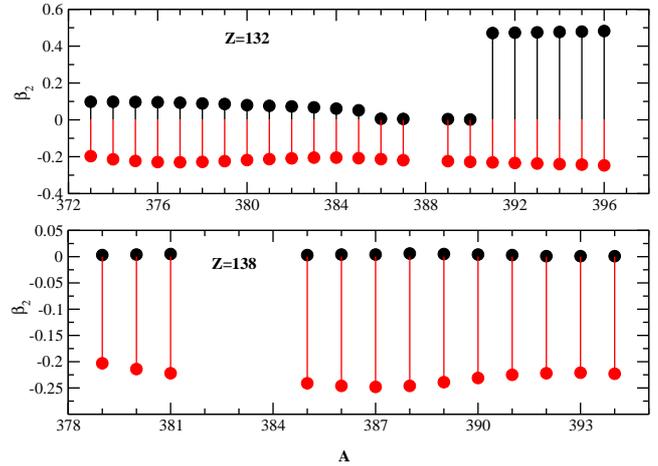}
}
\caption{(color online) Shape co-existence in $Z=132$ and $Z=138$ isotopic chain}
\label{oscillation}
\end{figure}

\subsection{Shape Coexistence}
One of the remarkable properties of nuclear quantum many body systems is
 its ability to minimize its energy by assuming different shapes at
 the cost of relatively small energy compared to its total binding energy.
 Generally, the nuclei having different binding energies correspond
 to their different shape configurations leads to the ground as well as intrinsic excited states. 
However, in certain cases it may happen that the binding energy of two different
 shape configurations may coincide or very close to each other and this
 phenomenon is known as shape coexistence~\cite{patra1993,sarazin2000,egido2004}. 
This phenomenon is more common in superheavy region giving rise 
to complex structures in these nuclei and thus enriching our
 understanding of the oscillations occurring between two or three
 existing shapes.
 In the isotopic chains discussed here in the paper, we have come across many
 examples where the ground and first excited binding energies are degenerate.
In the isotopic chain of $^{180-260}132$, we noticed the co-existence 
of shape (oblate-prolate, oblate-spherical) for $^{373-387}132$
 and $^{389-396}132$ isotopes as shown in Fig.~\ref{oscillation}. 
In present analysis, we consider a binding
 energy difference less or equal to 2 MeV for marking the shape co-existence.
 Due to this small binding energy difference the ground-state can change
 to low-lying excited state or vice verse by making a small change
 in the input parameters like the pairing energy.
The shape co-existence in 
nuclei indicates the competition between the different shape configurations 
differing from each other by a small amount in binding energy so as to 
acquire the ground state energy with maximum stability and the final shape
 could be a superposition of these low-lying bands. Further, in the 
isotopic chain of $^{180-260}138$, we noticed the shape co-existence (oblate-spherical)
 for $^{379-381}138$ and $^{385-394}138$ as shown in Fig.~\ref{oscillation}. 
Thus present analysis reveals that some of the nuclei of considered 
isotopic chain oscillate oblate-spherical as well as oblate-prolate and vice-verse.

\subsection{Density distribution}
\begin{figure}
\resizebox{0.47\textwidth}{!}{%
  \includegraphics{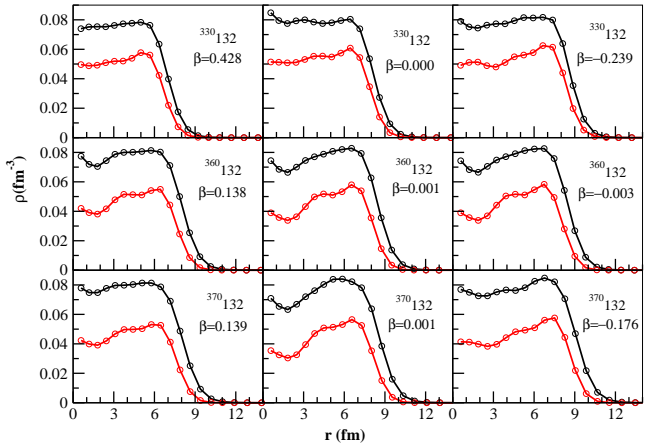}
}
\caption{(color online) Density profile for some selected nuclei 
on ground as well as intrinsic excited states.
The black line with circle represents the neutron 
density and proton density is shown by red line with red circle.}
\label{d132}
\end{figure}

\begin{figure}
\resizebox{0.47\textwidth}{!}{%
  \includegraphics{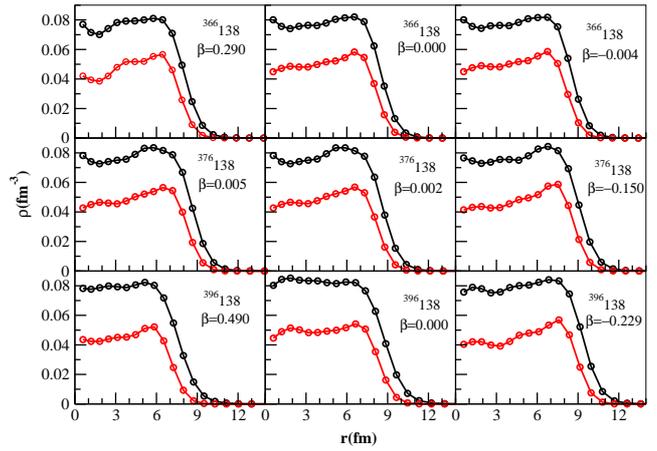}
}
\caption{(color online) Same as Fig.~\ref{d132} but for $^{366,376,396}138$.}
\label{d138}
\end{figure}

\begin{figure}
\resizebox{0.53\textwidth}{!}{%
  \includegraphics{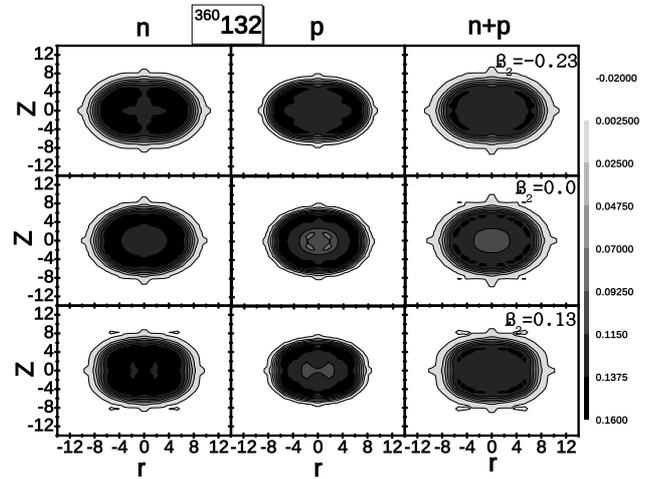}
}
\caption{Two dimensional neutron, proton and neutron plus 
proton density contours of $^{360}132$ nucleus for 
three different shape configurations.}
\label{cd132-360}
\end{figure}

\begin{figure}
\resizebox{0.53\textwidth}{!}{%
  \includegraphics{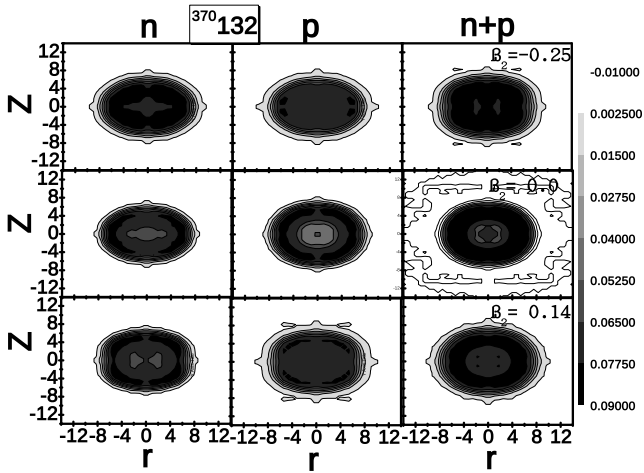}
}
\caption{Same as Fig.~\ref{d138} but for $^{370}132$.}
\label{cd132-370}
\end{figure}

Density distribution provides a detailed information regarding the 
distribution of nucleons for identifying central depletion in density, 
long tails and clusters in density plots. 
These features are known by bubble, halo and cluster
structures of the nuclei and may be observed in light to superheavy 
nuclei~\cite{whee,wilson1946,decharge2003,grasso2009,singh2013,sharma2006}.
Here, we have plotted the density profile for neutron, proton and total
matter (neutron plus proton) for some of the predicted closed shell 
nuclei~\cite{zhang2005,mbhuyan} within this framework as 
shown in Figs.~\ref{d132}, \ref{d138}.
Some of the nuclei for example; $^{360}132$ and $^{370}132$ show the depletion 
of central density on ground state as well as intrinsic excited states.
The strength of bubble shape is evaluated by calculating 
the depletion fraction~\cite{grasso2009,singh2013}.
There is no depletion of central density as such for 
$^{366}138$, $^{376}138$, $^{396}138$ systems.
Some nuclei such as $^{360}132$ and $^{370}132$ indicate a 
special kind of nucleon distribution.
In these cases, the centre is slightly bulgy and a considerably depletion 
afterward but again a big hump at mid of the centre and the surface.
To reveal such type of distribution and to gain an insight into the 
arrangement of nucleons, we make two-dimensional contour plots for 
$^{360}132$ and $^{370}132$ with three different 
shape configurations as given in Figs.~\ref{cd132-360} and \ref{cd132-370}. 
Figures~\ref{d132} and \ref{cd132-360} reflect that the hollow 
region at the centre is spread over the radius of $1-3$ fm.
This may suggest that these nuclei might have fullerene type 
structure and cluster of neutron and alpha-particle might be 
possibly within these types of nuclei. 
The full black contour refers to maximum density and full white ones
to zero density region.
It is apparent from figure~\ref{cd132-360} that the central portion 
of total matter density distribution in $^{360}132$ within spherical 
configuration is less dense than the peripheral region which 
can be interpreted as a thin gas of nucleons being surrounded by a 
thik sheath of nucleon (high density) giving rise to a bubble-type structure.
The individual neutron and proton density distributions also support 
the same bubble like structure within this shape configuration.
We witnessed a cluster type structure in total matter density 
distribution for oblate, spherical and prolate shape configurations. 
For the case of $^{370}132$ (Fig~\ref{cd132-370}), the two dimensional 
contour representation reveals that the total proton density distribution 
assumes a cluster shape for oblate and prolate configurations 
with $\beta_{2} =-0.25,0.14$ respectively. 
Whereas in case of spherical and prolate cases, the proton and total matter 
density distribution appears to be as bubble type, respectively. 
We noticed a semi-bubble like structure for the total nucleonic density 
distribution within the spherical case. 
The neutron density distribution plot for the oblate shape configuration
appears to be spindle shaped with prominent flaps/bulges.
Further, inspection reveals that the 
central part ($r=0$ fm) is considerably populated in proton density 
distribution but the depopulation is noticed at $r=1$ to $r=3$ fm 
and further a large population in proton density distribution 
beyond 3 fm is evident that goes to zero at the surface.  

\begin{table*}
\caption{Decay energies and half-lives of 
$\alpha$, $\beta$ and spontaneous fission for $Z=132$ isotopic chains.}
\renewcommand{\tabcolsep}{0.03cm}
\renewcommand{\arraystretch}{1.0}
\footnotesize\footnotesize
\begin{tabular}{cccccccccccccccc}
\hline\hline
Nuclei&$Q_\alpha^{RMF}$&$Q_\alpha^{FRDM}$&\multicolumn{6}{c}{$\log$($T_{1/2}^\alpha$)}&$\log$($T_{SF}$)&$Q_\beta^{RMF}$&$Q_\beta^{FRDM}$&\multicolumn{2}{c}{Fiset-Nix}&FRDM&Mode of \\
\cline{4-9} \cline{13-14}
&MeV&MeV&FRDM&VSS&GLDM&Brown&Royer&Ni et. al.&Ren-Xu&MeV&MeV&$\log$($T_{1/2}^\beta$)&$T_{1/2}^\beta$(sec)&$T_{1/2}^\beta$(sec)&decay\\
\hline
$^{312}132$&	17.70&&& 	-10.49& -10.33&	-9.30	&-9.70	&-10.82	&97.40		&11.07	&&-0.08	&0.84&&$\alpha$\\
$^{313}132$&	17.50&&&	-9.14&	-9.15&	-9.06	&-9.44	&-9.83	&83.03		&10.86	&&-0.53	&0.30&&$\alpha$\\
$^{314}132$&	17.14&&&	-9.68&	-9.56&	-8.62	&-8.93	&-10.13	&103.83		&10.82	&&-0.02	&0.96&&$\alpha$\\
$^{315}132$&	16.82&&&	-8.13&	-8.19&	-8.22	&-8.47	&-8.97	&87.03		&10.74	&&-0.50	&0.32&&$\alpha$\\
$^{316}132$&	16.68&&&	-8.98&	-8.91&	-8.04	&-8.28	&-9.53	&105.63		&10.5	&&0.06	&1.15&&$\alpha$\\
$^{317}132$&	16.51&&&	-7.65&	-7.74&	-7.81	&-8.03	&-8.56	&86.67		&10.21	&&-0.37	&0.43&&$\alpha$\\
$^{318}132$&	16.48&&&	-8.67&	-8.63&	-7.78	&-8.00	&-9.27	&102.89		&9.89	&&0.21	&1.63&&$\alpha$\\
$^{319}132$&	16.39&&&	-7.46&	-7.59&	-7.66	&-7.88	&-8.40	&82.01		&9.56	&&-0.20	&0.63&&$\alpha$\\
$^{320}132$&	16.34&&&	-8.45&	-8.45&	-7.59	&-7.82	&-9.08	&95.68		&8.58	&&0.57	&3.68&&$\alpha$\\
$^{321}132$&	16.30&&&	-7.32&	-7.48&	-7.54	&-7.77	&-8.28	&73.13		&8.42	&&0.12	&1.30&&$\alpha$\\
$^{322}132$&	16.05&&&	-7.98&	-8.02&	-7.19	&-7.39	&-8.68	&84.09		&8.17	&&0.69	&4.89&&$\alpha$\\
$^{323}132$&	15.92&&&	-6.70&	-6.90&	-7.01	&-7.19	&-7.75	&60.13		&8.08	&&0.22	&1.66&&$\alpha$\\
$^{324}132$&	14.38&&&	-5.01&	-5.10&	-4.70	&-4.48	&-6.14	&68.20		&7.95	&&0.76	&5.74&&$\alpha$\\
$^{325}132$&	14.28&&&	-3.75&	-4.00&	-4.54	&-4.30	&-5.23	&43.07		&7.63	&&0.36	&2.30&&$\alpha$\\
$^{326}132$&	14.19&16.01&-7.92&	-4.64&	-4.77&	-4.39	&-4.14	&-5.82	&48.10		&7.36	&8.70&0.95	&8.92&0.29&$\alpha$\\
$^{327}132$&	13.99&12.04&1.22 &	-3.17&	-3.47&	-4.05	&-3.76	&-4.74	&22.03		&7.07	&6.28&0.55	&3.56&0.81&$\alpha$\\
$^{328}132$&	13.95&11.57&1.38 &	-4.16&	-4.33&	-3.99	&-3.70	&-5.41	&23.85		&6.76	&4.88&1.16	&14.45&1.48&$\alpha$\\
$^{329}132$&	13.84&14.05&-3.30&	-2.87&	-3.20&	-3.80	&-3.50	&-4.48	&-2.92		&6.43	&5.73&0.78	&6.08&1.81&$\alpha$\\
$^{330}132$&	13.75&13.99&-4.23&	-3.75&	-3.96&	-3.64	&-3.33	&-5.06	&-4.46		&6.07	&4.30&1.42	&26.47&2.34&SF\\
$^{331}132$&	13.69&13.99&-3.02&	-2.56&	-2.92&	-3.54	&-3.22	&-4.21	&-31.71		&5.70	&5.16&1.08	&11.93&1.81&SF\\
$^{332}132$&	13.61&13.85&-3.96&	-3.46&	-3.71&	-3.40	&-3.07	&-4.82	&-36.76		&5.34	&3.61&1.73	&53.97&4.28&SF\\
$^{333}132$&	13.53&19.75&-12.10&	-2.22&	-2.62&	-3.26	&-2.92	&-3.92	&-64.27		&4.98	&10.32&1.40	&25.16&0.41&SF\\
$^{334}132$&	13.49&19.54&12.92&	-3.20&	-3.49&	-3.19	&-2.86	&-4.60	&-72.99		&4.65	&3.45&2.06	&15.38&8.71&SF\\
$^{335}132$&	13.38&13.88&-2.94&	-1.90&	-2.34&	-2.99	&-2.64	&-3.65	&-100.53	&4.35	&4.12&1.72	&52.64&15.82&SF\\
$^{336}132$&	13.36&19.20&-12.50&	-2.93&	-3.25&	-2.95	&-2.61	&-4.36	&-113.06	&4.14	&2.83&2.34	&16.89&15.69&SF\\
$^{337}132$&	13.14&14.40&-3.45&	-1.38&	-1.85&	-2.55	&-2.15	&-3.20	&-140.42	&4.01	&4.19&1.91	&81.90&11.55&SF\\
$^{338}132$&	12.90&14.04&-4.34&	-1.91&	-2.27&	-2.10	&-1.64	&-3.49	&-156.92	&3.91	&2.70&2.47	&96.07&18.05&SF\\
$^{339}132$&	12.72&13.33&-1.80&	-0.43&	-0.94&	-1.75	&-1.24	&-2.39	&-183.89	&3.71	&3.07&2.10     &24.50&$>$100&SF\\
$^{340}132$&	11.98&&&	0.31&	-0.11&	-0.24	&0.52	&-1.60	&-204.50	&3.66	&&2.63	&22.74&&SF\\
$^{341}132$&	11.76&&&	1.94&	1.38&	0.24	&1.07	&-0.37	&-230.87	&3.16	&&2.46	&89.71&&SF\\
$^{342}132$&	11.78&&&	0.82&	0.37&	0.19	&1.00	&-1.16	&-255.72	&2.77	&&3.25	&96.69&&SF\\
$^{343}132$&	12.13&&&	0.99&	0.41&	-0.56	&0.10	&-1.18	&-281.29	&1.99	&&3.47	&17.53&&SF\\
$^{344}132$&	12.86&&&	-1.82&	-2.28&	-2.02	&-1.64	&-3.41	&-310.54	&2.75	&&3.27	&76.72&&SF\\
$^{345}132$&	12.29&&&	0.60&	-0.02&	-0.89	&-0.32	&-1.52	&-335.10	&3.12	&&2.50	&13.69&&SF\\
$^{346}132$&	12.26&&&	-0.39&	-0.90&	-0.83	&-0.27	&-2.20	&-368.87	&2.90	&&3.16	&42.81&&SF\\
$^{347}132$&	12.06&&&	1.17&	0.52&	-0.41	&0.21	&-1.03	&-392.24	&2.73	&&2.80	&24.58&&SF\\
$^{348}132$&	11.85&&&	0.64&	0.09&	0.04	&0.72	&-1.32	&-430.67	&2.55	&&3.44	&81.75&&SF\\
$^{349}132$&	11.64&&&	2.25&	1.56&	0.50	&1.25	&-0.10	&-452.65	&2.35	&&3.12	&29.10&&SF\\
$^{350}132$&	11.55&&&	1.43&	0.84&	0.70	&1.47	&-0.64	&-495.88	&2.17	&&3.79	&4.75&&SF\\
$^{351}132$&	11.49&&&	2.66&	1.93&	0.84	&1.62	&0.24	&-516.28	&2.01	&&3.46	&59.16&&SF\\
$^{352}132$&	11.50&&&	1.56&	0.94&	0.82	&1.58	&-0.53	&-564.43	&1.82	&&4.16	&63.51&&SF\\
$^{353}132$&	11.52&&&	2.58&	1.82&	0.77	&0.21	&0.17	&-583.06	&1.62	&&3.90	&92.65&&SF\\
$^{354}132$&	11.66&&&	1.14&	0.48&	0.46	&1.12	&-0.89	&-636.28	&1.42	&&4.65	&74.39&&SF\\
$^{355}132$&	11.70&&&	2.10&	1.31&	0.37	&1.00	&-0.24	&-652.96	&1.25	&&4.40	&65.44&&SF\\
$^{356}132$&	11.71&&&	1.00&	0.31&	0.35	&0.96	&-1.01	&-711.35	&-1.05	&&5.21	&74.66&&SF\\
$^{357}132$&	11.81&&&	1.81&	0.99&	0.13	&0.68	&-0.48	&-725.91	&0.83	&&5.11	&38.01&&SF\\
$^{358}132$&	11.58&&&	1.35&	0.62&	0.64	&1.27	&-0.71	&-789.61	&0.89	&&5.50	&49.32&&SF\\
$^{359}132$&	11.21&&&	3.43&	2.57&	1.49	&2.26	&0.90	&-801.86	&0.73	&&5.32	&47.56&&SF\\
$^{360}132$&	10.83&&&	3.46&	2.69&	2.41	&3.33	&1.09	&-870.99	&0.57	&&6.18	&96.70&&SF\\
$^{361}132$&	10.45&&&	5.68&	4.77&	3.38	&4.46	&2.82	&-880.77	&0.38	&&6.20	&12.95&&SF\\
$^{362}132$&	10.11&&&	5.70&	4.88&	4.29	&5.52	&3.00	&-955.45	&0.14	&&7.61	&0.47&&SF\\
$^{363}132$&	10.07&&&	6.90&	5.95&	4.40	&5.63	&3.86	&-962.58	&-0.15	&&7.07	&93.35&&SF\\
$^{364}132$&	10.07&&&	5.83&	4.98&	4.40	&5.62	&3.11	&-1042.94	&-0.37	&&6.73	&41.85&&SF\\
$^{365}132$&	10.08&&&	6.86&	5.89&	4.37	&5.57	&3.83	&-1047.26	&-0.58	&&5.66	&7.56&&SF\\
$^{366}132$&	10.11&&&	5.70&	4.81&	4.29	&5.46	&3.00	&-1133.40	&-0.75	&&5.78	&49.94&&SF\\
$^{367}132$&	10.11&&&	6.77&	5.76&	4.29	&5.44	&3.75	&-1134.74	&-0.89	&&5.01	&2.06&&SF\\
$^{368}132$&	10.12&&&	5.67&	4.75&	4.26	&5.39	&2.97	&-1226.78	&-1.01	&&5.30	&85.03&&SF\\
\hline\hline
\end{tabular}
\label{tab5}
\end{table*}


\begin{table*}
\caption{Table V is continued.}
\renewcommand{\tabcolsep}{0.19cm}
\renewcommand{\arraystretch}{1.0}
\footnotesize
\begin{tabular}{cccccccccccc}
\hline\hline
Nuclei&$Q_\alpha$(cal.)&\multicolumn{5}{c}{$\log$($T_{1/2}^\alpha$)}&$\log$($T_{SF}$)&$Q_\beta$(cal.)&\multicolumn{2}{c}{Fiset-Nix}&Mode\\
\cline{3-7} \cline{10-11}
&MeV&VSS&GLDM&Brown&Royer&Ni et. al. &Ren-Xu&MeV&$\log$($T_{1/2}^\beta$)&$T_{1/2}^\beta$(sec)&of decay\\
\hline
$^{369}132$&	10.13&	6.70&	5.66&	4.23	&5.34	&3.69	&-1224.99	&-1.15	&4.57	&17.96&SF\\
$^{370}132$&	10.09&	5.77&	4.82&	4.34	&5.46	&3.06	&-1323.05	&-1.32	&4.82	&40.63&SF\\
$^{371}132$&	10.10&	6.80&	5.73&	4.32	&5.41	&3.78	&-1317.95	&-1.51	&4.06	&42.84&SF\\
$^{372}132$&	10.02&	6.00&	5.01&	4.54	&5.66	&3.26	&-1422.15	&-1.66	&4.38	&43.54&SF\\
$^{373}132$&	9.60&	8.50&	7.39&	5.75	&7.07	&5.23	&-1413.60	&-2.05	&3.45	&6.48&SF\\
$^{374}132$&	9.29&	8.56&	7.53&	6.69	&8.17	&5.45	&-1524.03	&-2.19	&3.81	&46.12&SF\\
$^{375}132$&	8.96&	10.89&	9.73&	7.75	&9.41	&7.27	&-1511.88	&-2.34	&3.17	&84.63&SF\\
$^{376}132$&	8.75&	10.67&	9.59&	8.46	&10.23	&7.24	&-1628.66	&-2.50	&3.53	&81.95&SF\\
$^{377}132$&	8.80&	11.53&	10.34&	8.29	&10.01	&7.82	&-1612.75	&-2.67	&2.89	&74.83&SF\\
$^{378}132$&	8.67&	11.00&	9.88&	8.73	&10.52	&7.52	&-1735.99	&-2.83	&3.26	&24.58&SF\\
$^{379}132$&	8.54&	12.60&	11.37&	9.19	&11.04	&8.73	&-1716.17	&-2.98	&2.65	&46.56&SF\\
$^{380}132$&	8.43&	12.01&	10.85&	9.58	&11.49	&8.39	&-1845.97	&-3.16	&3.02	&44.02&SF\\
$^{381}132$&	8.32&	13.55&	12.28&	9.99	&11.95	&9.54	&-1822.11	&-3.33	&2.40	&53.46&SF\\
$^{382}132$&	8.33&	12.44&	11.26&	9.95	&11.90	&8.76	&-1958.57	&-3.57	&2.75	&56.94&SF\\
$^{383}132$&	8.43&	13.07&	11.78&	9.58	&11.45	&9.13	&-1930.51	&-3.90	&2.05	&11.26&SF\\
$^{384}132$&	8.50&	11.71&	10.49&	9.33	&11.14	&8.13	&-2073.74	&-4.16	&2.40	&49.43&SF\\
$^{385}132$&	8.61&	12.31&	10.99&	8.94	&10.66	&8.48	&-2041.35	&-4.41	&1.76	&58.11&SF\\
$^{386}132$&	8.49&	11.75&	10.51&	9.37	&11.15	&8.17	&-2191.45	&-4.48	&2.23	&68.76&SF\\
$^{387}132$&	8.78&	11.61&	10.27&	8.36	&9.94	&7.88	&-2154.58	&-4.83	&1.55	&35.80&SF\\
$^{388}132$&	8.69&	10.91&	9.64&	8.67	&10.29	&7.45	&-2311.65	&-5.05	&1.95	&88.81&SF\\
$^{389}132$&	8.62&	12.27&	10.89&	8.91	&10.56	&8.45	&-2270.17	&-5.21	&1.38	&23.86&SF\\
$^{390}132$&	3.00&	61.12&	59.48&	50.81	&60.04	&50.30	&-2434.31	&-5.15	&1.91	&80.36&SF\\
$^{391}132$&	2.81&	66.23&	64.41&	54.21	&64.04	&54.51	&-2388.08	&-5.83	&1.11	&12.96&SF\\
$^{392}132$&	2.73&	66.99&	65.28&	55.74	&65.84	&55.32	&-2559.39	&-6.07	&1.52	&32.78&SF\\
\hline\hline
\end{tabular}
\label{tab6}
\end{table*}


\begin{table*}
\caption{Same as Table~\ref{tab5} but for $Z=138$ isotopic chain.}
\renewcommand{\tabcolsep}{0.20cm}
\renewcommand{\arraystretch}{1.0}
\footnotesize
\begin{tabular}{cccccccccccc}
\hline\hline
Nuclei&$Q_\alpha$(cal.)&\multicolumn{5}{c}{$\log$($T_{1/2}^\alpha$)}&$\log$($T_{SF}$)&$Q_\beta$(cal.)&\multicolumn{2}{c}{Fiset-Nix}&Mode\\
\cline{3-7} \cline{10-11}
&MeV&VSS&GLDM&Brown&Royer&Ni et. al.&Ren-Xu&MeV&$\log$($T_{1/2}^\beta$)&$T_{1/2}^\beta$(sec)&of decay\\
\hline
$^{318}138$&	 7.62&	18.17&	18.31&	14.50&	18.92&	13.77&	109.15	&	14.13&	-0.68&	0.21&$\beta$\\
$^{319}138$&	 7.10&	22.11&	22.08&	16.87&	21.75&	16.97&	98.02	&	13.83&	-1.12&	0.08&$\beta$\\
$^{320}138$&	 6.52&	24.65&	24.70&	19.84&	25.31&	19.30&	131.77	&	13.50&	-0.56&	0.27&$\beta$\\
$^{321}138$&	 6.05&	29.01&	28.89&	22.56&	28.55&	22.86&	117.23	&	13.15&	-0.99&	0.10&$\beta$\\
$^{322}138$&	 5.58&	31.65&	31.60&	25.61&	32.20&	25.27&	149.00	&	12.77&	-0.42&	0.38&$\beta$\\
$^{323}138$&	16.86&	-6.98&	-6.80&	-7.09&	-7.11&	-7.83&	131.31	&	12.37&	-0.84&	0.15&$\alpha$\\
$^{324}138$&	17.10&	-8.43&	-8.17&	-7.40&	-7.51&	-8.90&	160.94	&	11.90&	-0.24&	0.58&$\alpha$\\
$^{325}138$&	17.06&	-7.30&	-7.15&	-7.35&	-7.46&	-8.10&	140.38	&	11.82&	-0.72&	0.19&$\alpha$\\
$^{326}138$&	16.93&	-8.16&	-7.94&	-7.18&	-7.28&	-8.68&	167.70	&	11.57&	-0.17&	0.68&$\alpha$\\
$^{327}138$&	16.64&	-6.63&	-6.52&	-6.79&	-6.83&	-7.53&	144.51	&	11.35&	-0.62&	0.24&$\alpha$\\
$^{328}138$&	16.50&	-7.46&	-7.29&	-6.61&	-6.62&	-8.08&	169.36	&	11.17&	-0.08&	0.84&$\alpha$\\
$^{329}138$&	16.29&	-6.05&	-5.98&	-6.32&	-6.29&	-7.04&	143.81	&	11.02&	-0.54&	0.29&$\alpha$\\
$^{330}138$&	16.12&	-6.83&	-6.69&	-6.08&	-6.03&	-7.54&	166.01	&	10.75&	 0.02&	1.05&$\alpha$\\
$^{331}138$&	15.99&	-5.54&	-5.51&	-5.90&	-5.82&	-6.60&	138.35	&	10.47&	-0.41&	0.39&$\alpha$\\
$^{332}138$&	15.86&	-6.38&	-6.29&	-5.71&	-5.62&	-7.16&	157.75	&	10.21&	 0.15&	1.43&$\alpha$\\
$^{333}138$&	15.76&	-5.14&	-5.14&	-5.57&	-5.46&	-6.26&	128.23	&	9.97&	-0.28&	0.52&$\alpha$\\
$^{334}138$&	15.73&	-6.15&	-6.10&	-5.52&	-5.43&	-6.96&	144.67	&	9.74&	 0.27&	1.88&$\alpha$\\
$^{335}138$&	15.63&	-4.90&	-4.95&	-5.38&	-5.27&	-6.06&	113.52	&	9.51&	-0.16&	0.69&$\alpha$\\
$^{336}138$&	18.11&	-9.94&	-9.89&	-8.64&	-9.21& -10.19&	126.85	&	9.83&	 0.25&	1.80&$\alpha$\\
$^{337}138$&	15.61&	-4.87&	-4.95&	-5.35&	-5.27&	-6.03&	 94.31	&	9.06&	-0.04&	0.91&$\alpha$\\
$^{338}138$&	15.56&	-5.85&	-5.87&	-5.27&	-5.19&	-6.70&	104.38	&	8.80&	0.53&	3.39&$\alpha$\\
$^{339}138$&	15.52&	-4.71&	-4.82&	-5.21&	-5.14&	-5.90&	 70.68	&	8.50&	0.12&	1.32&$\alpha$\\
$^{340}138$&	15.28&	-5.33&	-5.40&	-4.85&	-4.72&	-6.27&	 77.33	&	8.21&	0.70&	5.06&$\alpha$\\
$^{341}138$&	15.07&	-3.88&	-4.03&	-4.53&	-4.35&	-5.19&	 42.71	&	7.93&	0.29&	1.96&$\alpha$\\
$^{342}138$&	14.82&	-4.46&	-4.57&	-4.14&	-3.90&	-5.53&	 45.79	&	7.63&	0.89&	7.70&$\alpha$\\
$^{343}138$&	14.70&	-3.17&	-3.36&	-3.94&	-3.68&	-4.58&	 10.46	&	7.24&	0.52&	3.29&$\alpha$\\
$^{344}138$&	14.63&	-4.09&	-4.24&	-3.83&	-3.56&	-5.21&	  9.84	&	6.95&	1.12&	13.08&$\alpha$\\
$^{345}138$&	14.27&	-2.30&	-2.54&	-3.24&	-2.86&	-3.85&	-25.97	&	6.93&	0.63&	 4.24&SF\\
$^{346}138$&	13.90&	-2.60&	-2.79&	-2.60&	-2.11&	-3.93&	-30.44	&	6.63&	1.23&	17.14&SF\\
$^{347}138$&	13.98&	-1.70&	-1.97&	-2.74&	-2.30&	-3.33&	-66.53	&	6.21&	0.90&	 7.85&SF\\
$^{348}138$&	14.06&	-2.94&	-3.16&	-2.88&	-2.48&	-4.22&	-75.00	&	5.88&	1.53&	33.58&SF\\
$^{349}138$&	13.96&	-1.66&	-1.96&	-2.70&	-2.29&	-3.30&	-111.14	&	5.60&	1.15&	14.00&SF\\
$^{350}138$&	13.47&	-1.66&	-1.93&	-1.83&	-1.25&	-3.13&	-123.73	&	5.59&	1.65&	44.68&SF\\
$^{351}138$&	21.29& -12.90& -13.13& -11.96& -13.45& -12.88&	-159.73	&	4.78&	1.52&	33.45&SF\\
$^{352}138$&	21.00& -13.64& -13.84& -11.69& -13.14& -13.35&	-176.59	&	5.96&	1.50&	31.60&SF\\
$^{353}138$&	14.25&	-2.26&	-2.63&	-3.20&	-2.95&	-3.81&	-212.23	&	5.42&	1.23&	16.98&SF\\
$^{354}138$&	14.19&	-3.21&	-3.53&	-3.10&	-2.85&	-4.45&	-233.50	&	5.23&	1.81&	65.25&SF\\
$^{355}138$&	14.12&	-1.99&	-2.39&	-2.98&	-2.72&	-3.58&	-268.58	&	5.03&	1.41&	25.71&SF\\
$^{356}138$&	14.04&	-2.89&	-3.25&	-2.84&	-2.57&	-4.19&	-294.38	&	4.82&	2.01&	 2.47&SF\\
$^{357}138$&	13.96&	-1.66&	-2.09&	-2.70&	-2.42&	-3.30&	-328.72	&	4.61&	1.62&	41.54&SF\\
$^{358}138$&	13.86&	-2.51&	-2.91&	-2.53&	-2.23&	-3.86&	-359.18	&	4.39&	2.23&	70.91&SF\\
$^{359}138$&	13.77&	-1.25&	-1.72&	-2.37&	-2.05&	-2.95&	-392.57	&	4.17&	1.86&	71.71&SF\\
$^{360}138$&	13.75&	-2.27&	-2.71&	-2.33&	-2.03&	-3.66&	-427.82	&	3.92&	2.50&	15.10&SF\\
$^{361}138$&	13.68&	-1.06&	-1.56&	-2.21&	-1.89&	-2.78&	-460.08	&	3.66&	2.16&	44.27&SF\\
$^{362}138$&	13.45&	-1.61&	-2.09&	-1.79&	-1.40&	-3.10&	-500.25	&	3.55&	2.73&	35.38&SF\\
$^{363}138$&	13.32&	-0.26&	-0.80&	-1.55&	-1.13&	-2.10&	-531.19	&	3.28&	2.41&	57.69&SF\\
$^{364}138$&	18.19& -10.05& -10.48&	-8.74&	-9.79& -10.29&	-576.40	&	3.03&	3.09&	26.59&SF\\
$^{365}138$&	13.17&	 0.09&	-0.49&	-1.27&	-0.82&	-1.81&	-605.83	&	2.86&	2.72&	24.89&SF\\
$^{366}138$&	13.05&	-0.70&	-1.24&	-1.04&	-0.56&	-2.32&	-656.21	&	2.75&	3.31&	24.64&SF\\
$^{367}138$&	12.96&   0.58&	-0.03&	-0.86&	-0.37&	-1.39&	-683.96	&	2.62&	2.92&	23.22&SF\\
$^{368}138$&	12.97&	-0.51&	-1.09&	-0.88&	-0.41&	-2.16&	-739.63	&	2.44&	3.57&	12.25&SF\\
$^{369}138$&	13.03&   0.41&	-0.23&	-1.00&	-0.56&	-1.53&	-765.51	&	2.24&	3.26&	 1.67&SF\\
$^{370}138$&	13.10&	-0.82&	-1.43&	-1.13&	-0.74&	-2.42&	-826.58	&	1.98&	4.01&	32.98&SF\\
$^{371}138$&	13.30&	-0.21&	-0.88&	-1.51&	-1.21&	-2.06&	-850.43	&	1.64&	3.90&	40.22&SF\\
$^{372}138$&	13.52&	-1.77&	-2.41&	-1.92&	-1.72&	-3.23&	-917.02	&	1.33&	4.81&	71.40&SF\\
$^{373}138$&	13.58&	-0.84&	-1.53&	-2.03&	-1.86&	-2.60&	-938.66	&	1.14&	4.59&	28.37&SF\\
$^{374}138$&	13.69&	-2.14&	-2.81&	-2.23&	-2.12&	-3.55&	-1010.89&	1.01&	5.31&	30.23&SF\\
\hline\hline
\end{tabular}
\label{tab7}
\end{table*}


\begin{table*}
\caption{Table VII is continued}
\renewcommand{\tabcolsep}{0.20cm}
\renewcommand{\arraystretch}{1.0}
\footnotesize
\begin{tabular}{cccccccccccc}
\hline\hline
Nuclei&$Q_\alpha$(cal.)&\multicolumn{5}{c}{$\log$($T_{1/2}^\alpha$)}&$\log$($T_{SF}$)&$Q_\beta$(cal.)&\multicolumn{2}{c}{Fiset-Nix}&Mode\\
\cline{3-7} \cline{10-11}
&MeV&VSS&GLDM&Brown&Royer&Ni et. al.&Ren-Xu&MeV&$\log$($T_{1/2}^\beta$)&$T_{1/2}^\beta$(sec)&of decay\\
\hline
$^{375}138$&	10.54&   7.26&   6.46&   4.64&	 6.12&	 4.31&	-1030.16&	1.59&	3.97&	70.88&SF\\
$^{376}138$&	10.12&   7.59&   6.81&   5.79&	 7.48&	 4.75&	-1108.13&	1.09&	5.18&	 7.04&SF\\
$^{377}138$&	 9.92&   9.35&   8.50&   6.36&	 8.16&	 6.09&	-1124.86&	0.76&	5.28&	26.15&SF\\
$^{378}138$&	12.47&   0.71&	-0.05&   0.12&	 0.64&  -1.12&	-1208.70&	0.31&	6.94&	20.87&SF\\
$^{379}138$&	16.40&	-6.23&	-6.96&	-6.47&  -7.30&  -7.20&	-1222.72&	0.10&	7.37&	93.17&SF\\
$^{380}138$&	16.24&	-7.03&	-7.75&	-6.25&  -7.05&  -7.71&	-1312.53&	-0.05&	8.28&	29.55&SF\\
$^{381}138$&	16.01&	-5.57&	-6.34&	-5.93&  -6.67&  -6.63&	-1323.69&	-0.22&	6.78&	43.46&SF\\
$^{382}138$&	15.94&	-6.52&	-7.27&	-5.83&  -6.57&  -7.28&	-1419.59&	-0.44&	6.55&	20.71&SF\\
$^{383}138$&	11.78&	3.58&	2.69&	1.61&	2.35&	1.17&	-1427.72&	-0.48&	5.95&	93.97&SF\\
$^{384}138$&	11.53&	3.21&	2.33&	2.18&	3.02&	1.01&	-1529.80&	-0.33&	6.89&	97.30&SF\\
$^{385}138$&	11.43&	4.56&	3.63&   2.41&	3.29&	2.00&	-1534.76&	-0.49&	5.92&	52.09&SF\\
$^{386}138$&	11.28&	3.92&	3.01&	2.77&	3.70&	1.62&	-1643.14&	-0.65&	6.03&	31.62&SF\\
$^{387}138$&	11.21&	5.19&	4.23&	2.94&	3.89&	2.55&	-1644.77&	-0.78&	5.26&	66.55&SF\\
$^{388}138$&	11.12&	4.39&	3.45&	3.16&	4.14&	2.03&	-1759.55&	-0.90&	5.52&	88.02&SF\\
$^{389}138$&	11.09&	5.55&	4.56&	3.23&	4.21&	2.85&	-1757.70&	-1.00&	4.85&	 7.15&SF\\
$^{390}138$&	11.01&	4.72&	3.74&	3.43&	4.43&	2.31&	-1878.98&	-1.10&	5.18&	45.38&SF\\
$^{391}138$&	10.81&	6.40&	5.37&	3.93&	5.02&	3.58&	-1873.50&	-1.18&	4.56&	59.33&SF\\
$^{392}138$&	10.65&	5.84&	4.82&	4.35&	5.51&	3.26&	-2001.38&	-1.26&	4.94&	61.70&SF\\
$^{393}138$&	10.43&	7.61&	6.54&	4.93&	6.19&	4.61&	-1992.14&	-1.35&	4.31&	29.21&SF\\
$^{394}138$&	10.27&	7.08&	6.02&	5.37&	6.70&	4.32&	-2126.72&	-1.46&	4.66&	49.79&SF\\
$^{395}138$&	3.82 & 52.22&  50.69&  41.66&  50.31&  42.64&	-2113.56&	-4.78&	1.59&	38.93&SF\\
$^{396}138$&	3.70 & 52.97&  51.48&  43.16&  52.10&  43.44&	-2254.94&	-3.08&	3.10&	59.17&SF\\
$^{397}138$&	3.68 & 54.35&  52.77&  43.42&  52.39&  44.46&	-2237.73&	-3.31&	2.44&	76.36&SF\\
$^{398}138$&	9.52 &  9.74&   8.59&   7.56&   9.28&   6.59&	-2386.00&	-3.47&	2.83&	82.99&SF\\
\hline\hline
\end{tabular}
\label{tab8}
\end{table*}

\subsection{Decay-energy and half-life}
In order to predict the modes of decay of the considered nuclides, the 
$\alpha$-decay, $\beta$-decay and SF half-lives shall be computed using 
various empirical formulas and their comparison of life-time shall 
predict the dominant mode of decay.
To analyze the dominant mode of decay(alpha), the alpha
decay half-lives are estimated using various empirical formulas such 
as Viola-Seaborg (VSS)~\cite{VJS66}, generalized liquid drop 
model (GLDM)~\cite{DR07}, Brown~\cite{B92}, Royer~\cite{R0},
 NI et al.~\cite{NDK08}. 
Spontaneous fission half-lives are computed using the 
semi-empirical formula of Ren and Xu~\cite{RX05} 
and made a comparison with alpha decay half lives to 
predict a possible decay mode of considered nuclides.
The beta decay half-lives are estimated using  empirical 
formulas of Fiset and Nix~\cite{FN72}.

\subsubsection{Alpha decay}  
A significant advancement has been made for estimating the alpha 
decay half-lives since the earliest formulation of Geiger and Nuttal~\cite{GN1122}. 
The calculation of $\alpha$ decay half life $T^{\alpha}_{\frac{1}{2}}$ 
requires the $Q_{\alpha}$ as input. 
The knowledge of $Q_{\alpha}$ of a nucleus gives a valuable information 
about its stability. 
The estimation of $Q_{\alpha}$ is done by knowing the binding 
energies of the parent and daughter nuclei and binding energy 
of the $^{4}$He. 
The binding energies are calculated using the versatile and powerful framework of relativistic mean-field theory. 
The $Q_{\alpha}$ energy is estimated using the relation
\begin{equation}
Q_{\alpha}(N,Z) = BE(N,Z) - BE(N-2,Z-2)-BE(2,2)
\end{equation}
Here, $BE(N,Z)$, $BE(N-2,Z-2)$, and $BE(2,2)$ are the binding energies of
 the parent, daughter and $^{4}$He (BE = 28.296 MeV~\cite{audi2003}) with neutron number N
 and proton number Z. With the even-even values available at hand, 
the alpha decay half-life of the isotopic chain under study is estimated
 by Viola-Seaborg semi-empirical relation
\begin{equation}
\log_{10}T^{\alpha}_{\frac{1}{2}}(sec) = \frac{aZ-b}{\sqrt{Q_{\alpha}}} -(cZ+d )+h_{\log}
\end{equation}
The values of the parameters a, b, c and d are taken from the recent 
modified parametrizations of Sobiczewski et al~\cite{SPC89}, which are
$a = 1.66175$, $b=8.5166$, $c=0.20$, $d=33.9069$. 
The $h_{\log}$ is the hindrance factor which takes into account the hindrance 
associated with odd proton and neutron numbers as given by Viola and Seaborg
\begin{equation}
h_{\log}= \left\{
\begin{array}{rl}
0 & even-even;\\
0.772 & odd-even;\\
1.066 &  even- odd; \\
1.114 & odd-odd;
\end{array} \right.
\end{equation}
The $Q_{\alpha}$ values obtained from RMF calculations are 
listed in the Tables~\ref{tab5} - \ref{tab8}. 
There are also several phenomenological formulas available 
in the literature for calculating the alpha decay half-lives. 
The semi-empirical formula proposed by Brown ~\cite{B92} for determining
 the half-life of superheavy nuclei is given by
\begin{equation}
\log_{10}T^{\alpha}_{\frac{1}{2}}(sec) = 9.54 (Z-2)^{0.6}/\sqrt{Q_{\alpha}} - 51.37
\end{equation}
where Z is the atomic number of parent nucleus and $Q_{\alpha}$ is in MeV.
Another formula proposed by Dasgupta-Schubert and Reyes~\cite{DR07} 
based on generalized liquid drop model and obtained by fitting the 
experimental half-lives for 373 alpha emitters for determining the 
half-lives of superheavy nuclei is given as
\begin{equation}
\log_{10}T^{\alpha}_{\frac{1}{2}}(sec) = a + bA^{1/6}Z^{1/2} + cZ/Q_{\alpha}^{1/2}  
\end{equation} 
The parameters a, b and c are given by
\begin{equation}
a,b,c= \left\{
\begin{array}{rl}
-25.31,-1.1629,1.5864 & even-even;\\
-26.65,-1.0859,1.5848 & even-odd;\\
-25.68,-1.1423,1.5920&  even- odd;\\ 
-29.48,-1.113,1.6971& odd-odd;
\end{array} \right.
\end{equation}

In Ref.~\cite{NDK08} Ni et. al. proposed a unified formula for 
determining the half-lives in alpha decay and cluster radioactivity.
The formula for alpha decay is written as
\begin{equation}
\log_{10}T^{\alpha}_{1/2}(sec) = 2a\sqrt{\mu} (Z-2) Q_{\alpha}^{-1/2} + b \sqrt{\mu}[2(Z-2)]^{-1/2} + c
\end{equation}
Where, a, b, c are the constants and $\mu$ is define as 4(A-4)/A.
Recently, Royer estimated the potential energy during $\alpha$ emission 
within liquid drop model including the proximity effects between 
$\alpha$ particle and the daughter nucleus and the $\alpha$ decay half-lives 
were calculated from the WKB barrier penetration probability 
analogous to asymmetric spontaneous fission. The theoretical predictions 
for half-life for heavy and superheavy nuclei by employing a 
fitting procedure to a set of 373 alpha emitters was developed by 
Royer~\cite{R0} with an RMS derivation of 0.42, given as
\begin{equation}
\log_{10}T_{1/2}(sec) = -26.06 -1.114A^{1/6}\sqrt{Z} + \frac{1.5837Z}{\sqrt{Q_{\alpha}}}
\end{equation}
where A and Z represent respectively the mass number and charge number 
of the parent nuclei and $Q_{\alpha}$ represents the energy released 
during the reaction. Assuming a similar dependence on A, Z and $Q_{\alpha}$, 
the above equation was reformulated for a subset of 131 even-even nuclei 
and a relation was obtained with a RMS derivation of only 0.285, given, as
\begin{equation}
\log_{10}T_{1/2}^{\alpha}(sec) = -25.31 -1.1629A^{1/6} \sqrt{Z} + \frac{1.5864Z}{\sqrt{Q_{\alpha}}}
\end{equation}
For a subset of 106 even-odd nuclei, the relation given by was further modified 
with an RMS derivation of 0.39, and is given as,
\begin{equation}
\log_{10}T_{1/2}^{\alpha}(sec) = -26.65 -1.0859A^{1/6} \sqrt{Z} + \frac{1.5848Z}{\sqrt{Q_{\alpha}}}
\end{equation}
A similar reformulation was performed for the equation for a 
subset of 86 odd-even nuclei and 50 odd-odd nuclei.

\subsubsection{Beta decay}
Beta decay is also a very important decay mode for proton-rich and
 neutron-rich nuclei. Fermi theory of $\beta$ decay involves 
electron-neutrino interaction, which describes the beta transition rates
 according to log(ft) values. We employed the empirical formula of
 Fiset and Nix ~\cite{FN72} for estimating the half-lives 
of the isotopic chain under study and is given as
\begin{equation}
T_{\beta} = 540 \times 10^{5.0} \frac{m_{e}^{5}}{\rho_{d.o.s.}(W_{\beta}^{6}-m_{e}^{6})}
\end{equation}
In an analogous way to $\alpha$ decay, we evaluate the $Q_\beta$
 value using the relation 
\begin{math}
Q_{\beta} = BE(Z+1,A)-B(Z,A)
\end{math}
and
\begin{math} 
W_{\beta} = Q_{\beta} + m_{e}
\end{math}
Here, $\rho_{d.o.s.}$ is the average density of states in the
 daughter nucleus ($e^{-A/290}\times$ number of states within 
1 MeV of ground state).

\subsubsection{Spontaneous Fission}
The determination of spontaneous half-lives helps in identifying the long
lived superheavy elements and mode of decay of heavy and superheavy nuclei. 
Several empirical formulas have been proposed by various authors from 
time to time for determining the spontaneous fission half-lives.
In our calculations, we employed the phenomenological formula proposed by 
Ren and Xu ~\cite{RX05} and is given by
\begin{small}
\begin{eqnarray}
 \log_{10}T^{\alpha}_{1/2}(sec)& = & 21.08 + C_{1} \frac{(Z-90-\nu)}{A}+C_{2}\frac{(Z-90-\nu)^{2}}{A}\nonumber\\
&+& C_{3}\frac{(Z-90-\nu)^{3}}{A} \nonumber\\ 
&+& C_{4} \frac{(Z-90-\nu)(N-Z-52)^{2}}{A} 
\end{eqnarray}
\end{small}
Where, Z, N, A represent the proton, neutron and mass number of parent nuclei.
C1, C2, C3, C4 are the empirical constants and $\nu$ is the seniority term 
which takes care of blocking effect of unpaired nucleons on the transfer 
of many nucleon pairs during the fission process.

\begin{figure}
\resizebox{0.52\textwidth}{!}{%
  \includegraphics{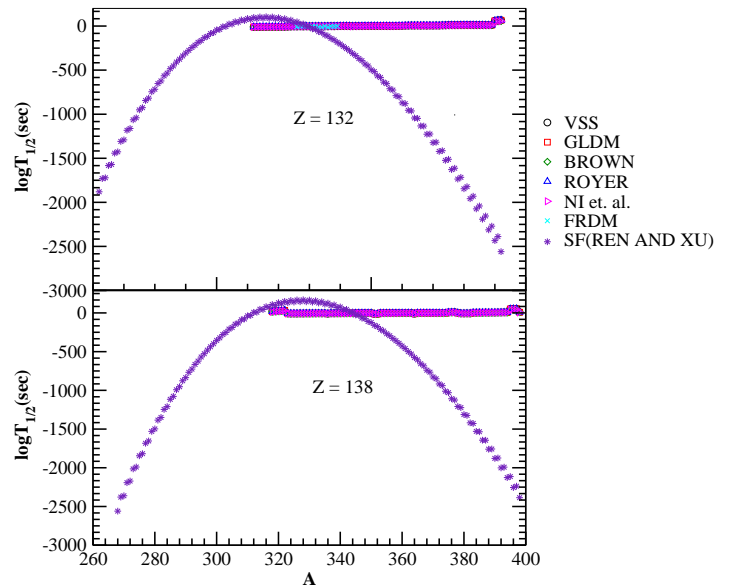}
}
\caption{(color online) Alpha decay and spontaneous fission half-lives of Z = 132, 138 
isotopic chain as a function of mass number.}
\label{alife}
\end{figure}

Our study on modes of decay highlights the range of isotopes which 
survive fission and thus decay through alpha emission.
Alpha and beta decay energies, $Q_\alpha$, $Q_\beta$ estimated by RMF 
binding energy is in quite agreement with FRDM data as given in Table~\ref{tab5}.
However, calculated half-lives by RMF do not match well with FRDM values.
This is why because $T_{\alpha} \propto 10^\frac{1}{\sqrt{Q_\alpha}}$ and 
$T_\beta \propto \frac{1}{{Q_\beta}^6}$ indicates that a small 
change in $Q_\alpha$ and $Q_\beta$ creates a big difference in 
$T_\alpha$ and $T_\beta$ as reflected in Tables~\ref{tab5} - \ref{tab8}.
The calculated alpha decay half-lives using VSS, GLDM, Brown, 
Royer and NI et al. are tabulated in Tables~\ref{tab5} - \ref{tab8} and a good 
agreement among them as well as with macro-microscopic data is noticed.
To check the possibility of $\beta$-decay empirical Fiset and Nix 
formula is employed to calculate the $\beta$-decay half-life for 
the considered isotopic chain and the results are given 
in Tables~\ref{tab5} - \ref{tab8}.
The beta decay half-lives are found to be very large than 
alpha decay as well as spontaneous fission half-lives and 
hence there is no possibility of mode of beta decay in 
present considered isotopes.
Further, SF half-lives is calculated and the values are 
framed in one of the columns of Tables~\ref{tab5} - \ref{tab8}.
Also, the alpha decay and SF half-lives for considered 
isotopic chain are plotted in Figure~\ref{alife}. 
After analyzing the concerned Tables and Figure, the analysis 
predicts that the isotopes of $Z=132$ with a mass range 
312 $\le A \le$ 329 survive the fission and may observed 
through alpha decay and those nuclei beyond $A>329$ do 
not survive the fission and hence completely undergo spontaneous fission.
For $Z=138$, the nuclides with a mass range 323 $\le A \le$ 344 
survive the fission and observed through alpha decay in the 
laboratory while the nuclei beyond $A>344$ do not 
survive fission and end with spontaneous fission.
In the first five isotopes $^{318-322}138$, beta decay is the
dominant mode of decay.
Therefore, the present study reveals that alpha decay 
and spontaneous fission are the principal modes of 
decay of majority of the considered nuclides with $beta$
decay as the principal mode of decay in first five isotopes
of $Z=138$.

\section{Summary}
We calculated the structural properties of $Z=132,138$ superheavy nuclei with a
range of neutron $N=180-260$ within axially deformed relativistic mean field theory.
The calculations are performed for prolate, oblate and spherical configurations 
in which prolate is suggested to be ground state.
The results produced by RMF are in good agreement with FRDM data.
Density distribution has been made to explain the special features of the 
nuclei such as bubble structure or halo structure.
Bubble structure is seen for some of the cases of the nuclei.
To make the clear presentation of nucleon distribution for some of 
the selected nuclei, the two-dimensional contour plot of density has 
been made by which cluster and bubble type structure is revealed.
Further, the predictions of possible modes of decay such as alpha decay, 
beta decay and spontaneous fission of the isotopes of $Z=132$ and $Z=138$ in 
the neutron range 180 $\le$ N $\le$ 260 have been done within self-consistent model.
The calculation of half-lives computed using Viola-Seaborg, GLDM, 
Brown, Royer and Ni et. al., show good agreement with each 
other as well as with macro-microscopic FRDM data.
All the physical observables calculated by RMF are found in 
good agreement with FRDM data. 
Also, the extensive study on beta decay half-lives and SF half-lives 
of the considered isotopic chain under investigations been made to 
identify the mode of the decay of these isotopes.
The study reveals that the isotopes of $Z=132$ that fall within the
mass range $312\le A \le 329$ undergoes alpha decay and those with mass 
number $A>329$ do not survive fission and hence completely undergoes
spontaneous fission. 
For $Z=138$, the alpha decay occurs within the isotopic mass chain 
318 $\le$ A $\le$ 344 and the isotopes beyond $A>344$ the mass range 
do not survive fission and end up with spontaneous fission.
The present analysis reveals that $\alpha$-decay and SF are the 
principal modes of decay in majority of the isotopes of superheavy
nuclei under study in addition to $\beta$ decay being the principal
mode of decay in $^{318-322}138$ isotopes.
Hence, we hope that present theoretical predictions on possible decay 
modes of $Z=132$, 138 superheavy nuclei might pave the way to help and 
guide the experimentalists in future for the synthesis of new superheavy isotopes.

\section{Acknowledgments}
One of the authors (MI) would like to acknowledge the 
hospitality provided by Institute of Physics (IOP), Bhubaneswar during the work.


\end{document}